\RequirePackage{fix-cm}
\documentclass[smallextended,envcountsect]{svjour3} 
\smartqed  
\usepackage{amsmath,amssymb,amsfonts}
\usepackage{algorithmic}
\usepackage{textcomp}
\usepackage{xcolor}
\usepackage{graphicx}
\usepackage{multirow}
\usepackage{makecell}
\usepackage{csquotes}
\usepackage{framed}
\usepackage{array}
\usepackage{hyperref}
\usepackage[capitalise]{cleveref}
\usepackage{dblfloatfix}    
\usepackage{natbib}
\usepackage{fontawesome}
\usepackage{soul}
\definecolor{greenncs}{rgb}{0.0, 0.62, 0.42}
\definecolor{airforceblue}{rgb}{0.36, 0.54, 0.66}
\colorlet{soulc}{greenncs!30}
\sethlcolor{soulc}
\newcolumntype{L}[1]{>{\raggedright\let\newline\\\arraybackslash\hspace{0pt}}m{#1}}
\newcolumntype{C}[1]{>{\centering\let\newline\\\arraybackslash\hspace{0pt}}m{#1}}
\newcolumntype{R}[1]{>{\raggedleft\let\newline\\\arraybackslash\hspace{0pt}}m{#1}}

\crefformat{footnote}{#2\footnotemark[#1]#3}
\begin{document}

\title{Identifying Self-Admitted Technical Debt in Issue Tracking Systems using Machine Learning
\thanks{This work was supported by ITEA3 and RVO under grant agreement No. 17038 VISDOM (\url{https://visdom-project.github.io/website}).}}
\titlerunning{Identifying SATD in Issue Tracking Systems using Machine Learning}        

\author{Yikun Li         \and
        Mohamed Soliman  \and
        Paris Avgeriou
}


\institute{Yikun Li \at
              Bernoulli Institute for Mathematics, Computer Science and Artificial Intelligence, University of Groningen, The Netherlands \\
              \email{yikun.li@rug.nl}
           \and
           Mohamed Soliman \at
              Bernoulli Institute for Mathematics, Computer Science and Artificial Intelligence, University of Groningen, The Netherlands \\
              \email{m.a.m.soliman@rug.nl}
           \and
           Paris Avgeriou \at
              Bernoulli Institute for Mathematics, Computer Science and Artificial Intelligence, University of Groningen, The Netherlands \\
              \email{p.avgeriou@rug.nl}
           \and
}

\date{Received: date / Accepted: date}

\maketitle

\begin{abstract}
Technical debt is a metaphor indicating sub-optimal solutions implemented for short-term benefits by sacrificing the long-term maintainability and evolvability of software. 
A special type of technical debt is explicitly admitted by software engineers (e.g. using a TODO comment); this is called Self-Admitted Technical Debt or SATD.
Most work on automatically identifying SATD focuses on source code comments.
In addition to source code comments, issue tracking systems have shown to be another rich source of SATD, 
but there are no approaches specifically for automatically identifying SATD in issues.
In this paper, we first create a training dataset by collecting and manually analyzing 4,200 issues (that break down to 23,180 sections of issues) from seven open-source projects (i.e., Camel, Chromium, Gerrit, Hadoop, HBase, Impala, and Thrift) using two popular issue tracking systems (i.e., Jira and Google Monorail).
We then propose and optimize an approach for automatically identifying SATD in issue tracking systems using machine learning.
Our findings indicate that: 1) our approach outperforms baseline approaches by a wide margin with regard to the F1-score; 2) transferring knowledge from suitable datasets can improve the predictive performance of our approach; 3) extracted SATD keywords are intuitive and potentially indicating types and indicators of SATD; 4) projects using different issue tracking systems have less common SATD keywords compared to projects using the same issue tracking system; 5) a small amount of training data is needed to achieve good accuracy.
\end{abstract}

\section{Introduction}
\label{sec:intro}

Technical debt (TD) is a metaphor reflecting the implementation or adoption of sub-optimal solutions to achieve short-term benefits while sacrificing the long-term maintainability and evolvability of software \citep{avgeriou_et_al:DR:2016:6693}.
TD is incurred either deliberately (e.g. to meet a deadline) or inadvertently (e.g. due to lack of domain knowledge) and tends to accumulate over time. 
If left unmanaged, the accumulation of TD can lead to critical issues in software maintainability and evolvability. 
To manage TD appropriately, a number of activities have been proposed, among which, \textit{identification} is the first step \citep{li2015systematic}.

Most TD identification approaches in literature use static source code analysis \citep{alves2016identification}. While this works well for identifying coding rule violations, it also has limitations. Most importantly, the scope of TD identification is limited to code-level issues such as code smells \citep{tufano2017and}. 
It does not consider other types of TD  (e.g. bad architecture decisions or requirements debt \citep{ernst2012role}), which are, in general, harder to detect from the source code directly.

Potdar and Shihab \citep{potdar2014exploratory} proposed a new approach to detect TD, focusing on the so-called \textit{Self-Admitted Technical Debt} (SATD): technical debt that is \textit{explicitly} admitted by software engineers in source code comments. The identification of SATD facilitates capturing TD items, which are harder to detect using source code analysis.
For instance, when developers choose to partially implement a requirement in order to deliver the software faster, they are likely to note that down in source code comments, e.g. by using a TODO comment \citep{potdar2014exploratory}.
Another example is the use of a technology (e.g. an architecture framework), not because it is the optimal solution, but because there is a business relation with the technology provider. Such examples of TD cannot be identified by analyzing source code.

Studies on identifying SATD, focus mostly on source code comments \citep{SIERRA201970}. 
Recently, issue tracking systems have shown to be another rich source for SATD, which acts as complementary to source code comments, i.e. it allows identifying SATD that is not admitted in source code comments \citep{9226330,bellomo2016got}.
In addition, issue tracking systems have the advantage of involving useful discussions between software engineers across the development activities (e.g. within requirements engineering, architecture design, and testing) \citep{merten2015requirements, 9226330}.
Therefore, there is high potential of finding comprehensive SATD content within issues; in contrast, source code comments hardly ever contain such discussions.
However, on the one hand, it is very time-consuming to manually identify SATD in issue tracking systems \citep{9226330}.
On the other hand, we lack approaches that automatically identify SATD in issue tracking systems.

In this paper, we aim at proposing and evaluating an approach for automatically identifying SATD in issue tracking systems. 
An issue usually consists of an issue summary, an issue description, and a number of comments from different developers.
We call each part of an issue (i.e. summary, description or comment) as \textit{issue section}.
We first collect 4,200 issues (that correspond to 23,180 issue sections) from seven large open-source projects from two ecosystems: Apache and Google.
We then manually analyze the issue sections in order to identify SATD based on our previously defined classification framework \citep{9226330}; this results in creating the biggest SATD dataset for issue tracking systems. 
Finally, based on this dataset for training, we experiment with several traditional machine learning and deep learning techniques to automatically identify SATD in issues within two popular issue tracking systems (i.e. Jira and Google Monorail). 
To optimize the machine learning outcome, we also explore different word embeddings, machine learning configurations (e.g. hyperparameter tuning), as well as transfer learning.

The main contributions of this paper are the following:

\begin{enumerate}
    \item \textbf{Contributing a rich dataset of self-admitted technical debt in issue tracking systems.}
    We collect 4,200 issues (that contain 3,277 SATD issue sections out of 23,180 issue sections) from seven open-source projects using two issue tracking systems.
    The dataset includes annotations regarding the type and indicator of each SATD issue section.
    We make our dataset publicly available\footnote{\label{l:data}\url{https://github.com/yikun-li/satd-issue-tracker-data}} to encourage future research in this area.

    \item \textbf{Comparing different machine learning techniques and optimizing the best approach to identify SATD in issue tracking systems.}
    We compare the F1-score of different machine learning approaches identifying SATD in issues and find out that Text CNN \citep{kim2014convolutional} outperforms others.
    We then further investigate imbalanced data handling strategies, word embedding techniques, and hyperparameter tuning, and optimize the CNN-based approach to accurately identify SATD issue sections in issue tracking systems.
    Moreover, we conduct extensive experiments to evaluate the effectiveness of transferring knowledge gained from other datasets.
    
    \item \textbf{Extracting the most informative SATD keywords and comparing keywords from different projects and sources.}
    We summarize and present a list of the most informative SATD keywords and we find that these keywords are intuitive and can potentially indicate types and indicators of SATD.
    Besides, 
    we show that projects using different issue tracking systems have less common keywords compared to projects using the same issue tracking system.
    Moreover, we find source code comments and issue tracking systems have some common SATD keywords.

    \item \textbf{Evaluating generalizability of our CNN-based approach.}
    We conduct experiments to evaluate the generalizability of our approach across projects and issue tracking systems.
    
    \item \textbf{Exploring the amount of data necessary for training the model.}
    We find that only a small amount of training data is needed to achieve good accuracy. 
\end{enumerate}

The remainder of the paper is organized as follows.
We begin by discussing some related work in \cref{sec:related}.
We then elaborate on the case study design in \cref{sec:approach}.
Subsequently, we present and discuss the results in \cref{sec:results} and \cref{sec:discussion} respectively.
In \cref{sec:validity}, threats to validity are discussed.
Finally, we present our conclusions in \cref{sec:conclusion}.

\section{Related Work}
\label{sec:related}

In this paper, we work on an approach to identify SATD in issue tracking systems.
Therefore, we divide the related work into two parts: work related to SATD in general and work related to SATD in issue tracking systems.

\subsection{Self-Admitted Technical Debt}

\cite{potdar2014exploratory} investigated source code comments that indicate technical debt items and named this phenomenon \emph{self-admitted technical debt}.
They manually analyzed 101,762 source code comments from four open-source projects (i.e., Eclipse, Chromium, Apache HTTP Server, and ArgoUML) to identify SATD comments.
They found that SATD comments are widely spread in projects: 2.4\% to 31.0\% of files contain SATD comments.
Moreover, they identified and summarized 62 keyword phrases that indicate SATD, such as \textit{ugly}, \textit{temporary solution}, and \textit{this doesn't look right}.
In a follow-up study, \cite{maldonado2015detecting} identified SATD by reading through 33,093 comments from five open-source project and manually classifying them into different types.
They found that source code comments indicate five types of SATD: design, defect, documentation, requirement, and test debt.
Besides, they observed that the majority of SATD is design debt as 42\% to 84\% of all identified SATD comments indicate design debt.

Following up from \cite{maldonado2015detecting}, to accurately and automatically identify SATD in source code comments, \cite{da2017using} analyzed 29,473 source code comments from ten open-source projects and trained a maximum entropy classifier on the analyzed data.
The results showed that design and requirement debt can be identified with the average F1-score of 0.620 and 0.402 respectively.
Additionally, they found that training on a small subset of comments can achieve 80\% and 90\% of the best classification accuracy.

Subsequently, there was work on improving the accuracy of SATD identification in source code comments.
\cite{de2016investigating} investigated the effectiveness of Contextualized Vocabulary Model for identifying SATD in code comments.
\cite{liu2018satd} proposed an approach based on text-mining to accurately and automatically detect SATD in source code comments by utilizing feature selection and combing sub-classifiers.
The average F1-score achieved by their approach was improved from 0.576 to 0.737, compared to the work by \cite{da2017using}.
Most recently, \cite{ren2019neural} proposed a Convolutional Neural Network based approach for SATD identification in source code comments and showed that their approach outperformed previous methods.

Apart from SATD identification, there has been work related to measurement and repayment of SATD.
\cite{wehaibi2016examining} investigated the relation between SATD and software quality by analyzing five open-source projects, namely Chromium, Hadoop, Spark, Cassandra, and Tomcat.
The findings indicated that SATD changes (modifications on files containing SATD comments) incur fewer defects compared to non-SATD changes, while SATD changes are more complex and difficult to perform than non-SATD changes.
\cite{kamei2016using} found that about 42\% to 44\% of SATD sections incur positive technical debt interest.
\cite{zampetti2018self} studied how SATD is resolved in five open-source projects, namely Camel, Gerrit, Hadoop, Log4j, and Tomcat.
They found that between 20\% and 50\% of SATD comments are removed by accident (without addressing the SATD), while most of the repayment activities require complex source code changes.

\subsection{SATD in Issue Tracking Systems}

SATD in issue tracking systems is relatively unexplored: there are only four studies on the identification and repayment of SATD in issue trackers.
\cite{bellomo2016got} explored the existence of SATD in four issue tracking systems from two government projects and two open-source projects.
They analyzed 1,264 issues and annotated 109 issues as SATD issues.
The results showed that technical debt is indeed declared and discussed in issue tracking systems.
Subsequently, \cite{dai2017detecting} identified 331 SATD issues from 8,149 issues by reading through the issue summaries and descriptions.
They then trained a Naive Bayes classifier to automatically classify issues as SATD issues or non-SATD issues and extracted unigram keywords that indicate technical debt.
The third study by \cite{9226330} is our own previous work, where we investigated the identification and repayment of SATD in issues from two open-source projects.
We annotated issues on the sentence level, instead of treating a whole issue as SATD or not, in order to have better accuracy.
We then presented types of SATD, the points of time when SATD was identified and reported, and how SATD was eventually resolved.
Lastly, \cite{xavier2020beyond} studied a sample of 286 SATD issues and found 29\% of SATD in issues can be tracked to source code comments.

In this article, we analyze issues on a more fine-grained level compared to three of the related studies \citep{bellomo2016got,dai2017detecting,xavier2020beyond} as they both treated a whole issue as a single technical debt statement. In contrast, we look at issue sections (i.e., individual issue summaries, descriptions, or comments) and potentially annotate them as SATD issue sections.
As compared to the third study \citep{9226330}, we analyze about eight times more issues (4,200 issues versus 500 issues) for machine learning training.
Besides, we propose a deep learning approach to accurately identify SATD in issues and compare the accuracy with other traditional machine learning methods (see RQ1).
Moreover, we extract and summarize unigram to five-gram keywords, compared to only unigram keywords extracted from issue descriptions by \cite{dai2017detecting} (see RQ2).
Furthermore, we investigate the generalization of our approach (see RQ3) and explore the amount of data needed for training the machine learning model (see RQ4).
RQ3 and RQ4 were not investigated before by other researchers.
Finally, this is the first work discussing the differences between SATD from different sources (i.e., source code comments and issue tracking systems).

\section{Study Design}
\label{sec:approach}

We follow the guidelines for case study research proposed by \cite{runeson2012case} to design and conduct the study.
The goal of the study, formulated according to the Goal-Question-Metric \citep{Solingen:02} template, is to ``\textit{\textbf{analyze} issues in issue tracking systems \textbf{for the purpose of} automatically identifying self-admitted technical debt within issues \textbf{with respect to} accuracy, explainability, and generalizability \textbf{from the point of} developers \textbf{in the context of} open-source software}''.
This goal is refined into four research questions (RQs), which consider \textit{accuracy} (RQ1 and RQ4), \textit{explainability} (RQ2) and \textit{generalizability} (RQ3).
The RQs and their motivations are explained below.

\begin{itemize}
    \item \textbf{(RQ1)} \textit{How to accurately identify self-admitted technical debt in issue tracking systems?}
    This research question focuses on the \emph{accuracy} of SATD automatic identification, and is further refined into three sub-questions:
    
    \begin{itemize}
        \item \textbf{(RQ1.1)} \textit{Which algorithms have the best accuracy to capture self-admitted technical debt in issue tracking systems?}
        Since we are aiming at accurately identifying SATD within issues, we need to compare the accuracy of different approaches to choose the best one.
        
        \item \textbf{(RQ1.2)} \textit{How to improve accuracy of machine learning model?}
        To optimize the accuracy of identifying SATD in issue tracking systems, we investigate word embedding refinement, imbalanced data handling strategies, and hyperparameters tuning. 
        
        \item \textbf{(RQ1.3)} \textit{How can transfer learning improve the accuracy of identifying self-admitted technical debt in issue tracking systems?}
        Transfer learning focuses on using knowledge gained while solving one task to address a different but related task.
        Therefore, we can study the influence of leveraging external datasets (from source code comments) on our SATD detector using transfer learning technique.
    \end{itemize}
    
    \item \textbf{(RQ2)} \textit{Which keywords are the most informative to identify self-admitted technical debt in issue tracking systems?}
    By extracting keywords of technical debt statements, we can understand better how developers declare technical debt in issue tracking systems.
    The summarized keywords are also helpful to developers for understanding and identifying SATD within issues.
    Overall, understanding these keywords allow us to \emph{explain} how the classifier works.
    
    \item \textbf{(RQ3)} \textit{How generic is the classification approach among projects and issue tracking systems?}
    Different projects use different issue tracking systems (e.g. Jira and Google Monorail) and are maintained by different communities (e.g. Apache and Google). 
    Thus, we need to evaluate 
    how far our results can be applicable to different projects and issue tracking systems.
    This research question is concerned with the \emph{generalizability} of machine learning approaches.
    
    \item \textbf{(RQ4)} \textit{How much data is needed for training the machine learning model to accurately identify self-admitted technical debt in issues?}
    Intuitively, training a machine learning classifier on a bigger dataset leads to better accuracy. 
    However, manually annotating SATD in issue tracking systems is a time-consuming task. 
    Therefore, we ask RQ4 to determine the most suitable size for a training dataset, which can achieve the best classification \emph{accuracy} with a minimum amount of effort.
\end{itemize}

\subsection{Approach Overview}
\label{sec:approach_overview}

\begin{figure*}[t]
  \centering
  \includegraphics[width=\linewidth]{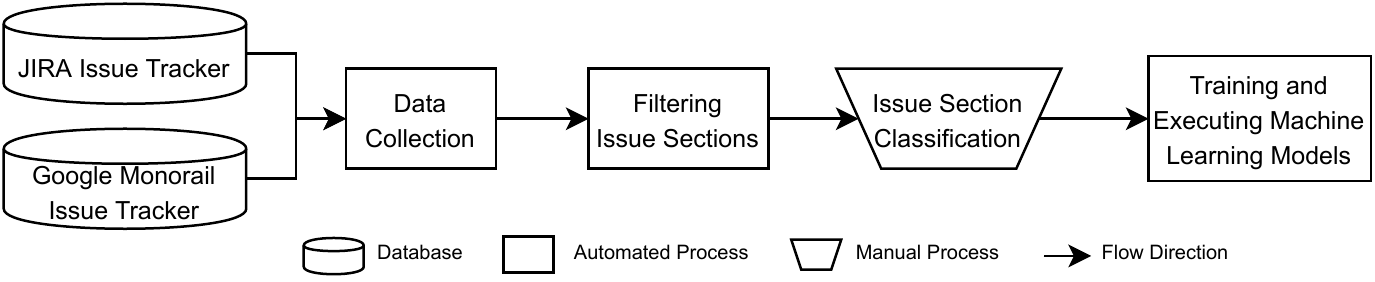}
  \caption{The framework of our approach.}
  \label{f:framework}
  \vspace{-3mm}
\end{figure*}

\cref{f:framework} presents an overview of our approach.
The first step is to collect issue sections from projects that use Jira and Google Monorail.
Subsequently, collected issue sections are filtered to remove impertinent data. 
Then, issue sections are manually classified regarding their SATD.
Finally, the machine learning models are trained on the manually classified dataset and executed on the whole dataset. 
Each of these steps is elaborated in the following sub-sections; the last step is the most complex, so it is explained in more depth.

\subsection{Data Collection}

To automatically identify SATD in issues, we will use supervised machine learning classifiers \citep{shalev2014understanding}, which depend on training data that are manually collected and classified.
Thus, we first need to manually collect and analyze issues from different issue tracking systems.

Two of the most mainstream issue tracking systems are Jira and Google Monorail.
Furthermore, projects, such as Camel, Chromium, Gerrit, and Hadoop, in the Apache and Google ecosystems have been commonly used in other SATD studies \citep{potdar2014exploratory,wehaibi2016examining,zampetti2018self}.
Finally, projects in the Apache and Google ecosystems are of high quality and supported by mature communities \citep{tom18}.
Thus, we looked into projects in the Apache and Google ecosystems.
Projects from the two ecosystems use two different mainstream issue tracking systems, namely: the Jira issue tracking system\footnote{\url{https://www.atlassian.com/software/jira}} and the Google Monorail issue tracking system\footnote{\url{https://bugs.chromium.org/}}.
To select projects pertinent to our study goal, we set the following criteria:

\begin{enumerate}
    \item Both the issue tracking system and the source code repository are publicly available.
    
    \item They have at least 200,000 source lines of code (SLOC) and 5,000 issues in the issue tracking systems.
    This is to ensure sufficient complexity.
\end{enumerate}

\begin{table}[t]
\caption{Details of chosen projects.}
\label{tb:projects}
\begin{center}
\resizebox{\columnwidth}{!}{
\def\arraystretch{1.2}
\begin{tabular}{@{\extracolsep{4pt}}C{1.65cm}C{1cm}C{3.2cm}C{1.1cm}C{1.3cm}C{1.6cm}C{1.2cm}C{1.2cm}@{}}
\hline
\multirow{2}{*}{\textbf{Project}} & \multicolumn{4}{c}{\textbf{Project Details}} & \multicolumn{3}{c}{\textbf{Classification Details}} \\
\cline{2-5}
\cline{6-8}
 & Issue Tracker & Languages & SLOC & \# Issues & \# Analyzed Sections & \# SATD Sections & \% SATD Sections \\
\hline
\textbf{Camel} & Jira & Java & 1,525k & 14,411 & 2,792 & 377 & 13.5\% \\
\textbf{Chromium} & Google & C++, C, and JavaScript & 22,472k & 1,079,511 & 3,435 & 264 & 6.7\% \\
\textbf{Gerrit} & Google & Java & 455k & 12,711 & 2,812 & 195 & 6.9\% \\
\textbf{Hadoop} & Jira & Java & 3,409k & 16,808 & 4,515 & 831 & 18.4\% \\
\textbf{HBase} & Jira & Java & 912k & 24,342 & 4,936 & 688 & 13.9\% \\
\textbf{Impala} & Jira & C++, Java, and Python & 640k & 9,733 & 1,934 & 355 & 18.4\% \\
\textbf{Thrift} & Jira & C++, Java, and C & 294k & 5,196 & 2,756 & 567 & 20.6\% \\
\hline
\textbf{Average} & & & & & 3,311 & 468 & 14.1\% \\
\textbf{Total} & & & & & 23,180 & 3,277 & \\
\hline
\end{tabular}
}
\end{center}
\vspace{-5mm}
\end{table}

There are over 400 projects in the Apache and Google ecosystems.
However, training machine learning models requires manually analyzing issues to create the training dataset. Therefore, we randomly selected a sample of seven projects to be used for training and testing the machine learning models. This number of projects is similar to other SATD studies, which analyzed from four to ten projects \citep{potdar2014exploratory,wehaibi2016examining,da2017using}.
The details of the chosen projects are shown in Table~\ref{tb:projects}.
We analyzed the latest released versions on May 7, 2020.
The number of source lines of code (SLOC) is calculated using the LOC tool\footnote{\url{https://github.com/cgag/loc}}.
It should be noted that although Chromium has significantly more issues than other projects, we found through manual inspection that the issues of Chromium are similar to the issues of other projects (and esp. Gerrit) with respect to the creation and discussion of issues. This is further supported by the fact that the same percentage (approx. 7\%) of issue sections from Chromium and Gerrit were classified as SATD sections; thus the issues of Chromium and Gerrit are maintained similarly.

\subsection{Filtering Issue Sections}

In issue tracking systems, in addition to the comments submitted by software developers, some comments are automatically generated by bots (e.g., the Jenkins bot generates comments to report the results of the Jenkins build).
Therefore, we filtered out comments that are automatically generated by bots.
Specifically, we first obtain the 100 most active users by ordering the number of comments submitted per user. 
Then we identify the bots' usernames (such as \textit{Hadoop QA} and \textit{Hudson}) by checking comments submitted by these 100 most active users.
After that, we removed all comments posted by bot users according to the list of bots' usernames.

Additionally, software developers sometimes attach source code in issues for various reasons.
Because we focus on the SATD in issues rather than in source code, we removed source code in issues using a set of regular expressions\cref{l:data}.

\subsection{Issue Section Classification}

Before training machine learning models to automatically identify SATD in issues, we need to inspect the collected sections within issues and manually classify them.
Since software developers might discuss several types of SATD in the same issue, treating a whole issue as a single type of technical debt may be inaccurate.
For example, software developers discussed both code debt and test debt in issue \emph{HADOOP-6730}\footnote{\url{ https://jira.apache.org/jira/browse/HADOOP-6730}}.
To accurately identify SATD in issue tracking systems, similarly to our previous study \citep{9226330}, we treat issue summaries, descriptions, and comments as separate sections, and classify each section individually.

\begin{table}[thpb]
\caption{Types and indicators of self-admitted technical debt.}
\label{tb:statistics_type_projects}
\begin{center}
\resizebox{\columnwidth}{!}{
\def\arraystretch{1.2}
\begin{tabular}{@{\extracolsep{4pt}}L{2
cm}L{3.6cm}L{5.8cm}C{0.5cm}C{0.6cm}C{0.6cm}@{}@{}}
\hline
\textbf{Type} & \textbf{Indicator} & \textbf{Definition} & \textbf{\#} & \textbf{\#} & \textbf{\%} \\
\hline
\multirow{3}{*}{\makecell[l]{Architecture \\debt}} & Violation of modularity & Because shortcuts were taken, multiple modules became inter-dependent, while they should be independent. & 46 & \multirow{3}{*}{87} & \multirow{3}{*}{2.7} \\
\cline{2-4}
 & Using obsolete technology & Architecturally-significant technology has become obsolete. & 41 &  \\
\hline
\multirow{3}{*}{Build debt} & Over- or under-declared dependencies & Under-declared dependencies: dependencies in upstream libraries are not declared and rely on dependencies in lower level libraries. \newline Over-declared dependencies: unneeded dependencies are declared. & 25 & \multirow{3}{*}{64} & \multirow{3}{*}{2.0} \\
\cline{2-4}
 & Poor deployment practice & The quality of deployment is low that compile flags or build targets are not well organized. & 39 &  \\
\hline
\multirow{10}{*}{Code debt} & Complex code & Code has accidental complexity and requires extra refactoring action to reduce this complexity. & 30 & \multirow{10}{*}{1246} & \multirow{10}{*}{38.0} \\
\cline{2-4}
 & Dead code & Code is no longer used and needs to be removed. & 121 &  \\
\cline{2-4}
 & Duplicated code & Code that occurs more than once instead of as a single reusable function. & 40 &  \\
\cline{2-4}
 & Low-quality code & Code quality is low, for example because it is unreadable, inconsistent, or violating coding conventions. & 856 &  \\
\cline{2-4}
 & Multi-thread correctness & Thread-safe code is not correct and may potentially result in synchronization problems or efficiency problems. & 40 &  \\
\cline{2-4}
 & Slow algorithm & A non-optimal algorithm is utilized that runs slowly. & 159 &  \\
\hline
\multirow{1}{*}{Defect debt} & Uncorrected known defects & Defects are found by developers but ignored or deferred to be fixed. & 25 & 25 & 0.8 \\
\hline
\multirow{1}{*}{Design debt} & Non-optimal decisions & Non-optimal design decisions are adopted. & 935 & 935 & 28.5 \\
\hline
\multirow{4}{*}{\makecell[l]{Documentation \\debt}} & Low-quality documentation & The documentation has been updated reflecting the changes in the system, but quality of updated documentation is low. & 342 & \multirow{4}{*}{486} & \multirow{4}{*}{14.8} \\
\cline{2-4}
 & Outdated documentation & A function or class is added, removed, or modified in the system, but the documentation has not been updated to reflect the change. & 144 &  \\
\hline
\multirow{4}{*}{\makecell[l]{Requirement \\debt}} & Requirements partially implemented & Requirements are implemented, but some are not fully implemented. & 67 & \multirow{4}{*}{96} & \multirow{4}{*}{2.9} \\
\cline{2-4}
 & Non-functional requirements not being fully satisfied & Non-functional requirements (e.g. availability, capacity, concurrency, extensibility), as described by scenarios, are not fully satisfied. & 29 &  \\
\hline
\multirow{6}{*}{Test debt} & Expensive tests & Tests are expensive, resulting in slowing down testing activities. Extra refactoring actions are needed to simplify tests. & 28 & \multirow{6}{*}{338} & \multirow{6}{*}{10.3} \\
\cline{2-4}
 & Flaky tests & Tests fail or pass intermittently for the same configuration. & 83 &  \\
\cline{2-4}
 & Lack of tests & A function is added, but no tests are added to cover the new function. & 158 &  \\
\cline{2-4}
 & Low coverage & Only part of the source code is executed during testing. & 69 &  \\
\hline
\end{tabular}
}
\end{center}
\end{table}

Since manual classification is extremely time consuming, we are not able to analyze all issues.
Thus, we calculated the size of the statistically significant sample based on the total number of issues of each project with a confidence level of 95 percent and a confidence interval of 5 percent.
We found that the sizes of statistically significant samples range from 358 to 384.
Therefore, we randomly selected 600 issues from each project for analysis to ensure each sample size is greater than the statistically significant sample size.
We then decomposed issues into sections (summaries, descriptions, and comments) for manual classification.
Because each issue contains one summary, one description, and several comments, the number of decomposed sections per issue varies.
This step resulted in 23,180 issue sections for analysis (see \cref{tb:projects}).

We used an open-source text annotation tool (Doccano\footnote{\url{https://github.com/doccano/doccano}}) to annotate sections using an existing classification framework from our previous work \citep{9226330}. 
This classification framework contains several types and indicators of SATD (see \cref{tb:statistics_type_projects}), and was based on the original framework by \citep{alves2014towards}.
Based on this classification framework, if a section matches an indicator (i.e. it indicates a certain type of SATD), we classify the section as the corresponding type. 
Classifying SATD into these types allows for a comparison between classifying SATD in issue tracking systems versus SATD in source code comments (see \cref{sec:discussion}).

Subsequently, all issue sections were divided into four subsets and assigned randomly for analysis to four independent researchers, who are different from the authors of this paper. We selected the four independent researchers using convenience sampling; this resulted in an average of 5795 issue sections per independent researcher.
To prepare the independent researchers for this task, we gave them a tutorial about SATD and explained to them the definitions of the different types and indicators of SATD in the classification framework (shown in \cref{tb:statistics_type_projects}).
Moreover, each type of SATD was supported with examples from multiple projects in issues. 
This facilitated understanding each type of technical debt.

The next step for each independent researcher was to analyze and manually classify 200 issue sections (according to the classification framework in \cref{f:framework}).
These results were compared and discussed with the first author to ensure that the annotations of each independent researcher align with the definitions of the types of SATD in the classification framework.
This process was repeated three times for each independent researcher to ensure a better and uniform understanding of the types of SATD.

Finally, all four independent researchers finished classifying the issue sections in his/her subset; thereafter we measured the level of agreement between the classifications of independent researchers and the first author using Cohen's kappa coefficient \citep{fleiss1981measurement}, which is commonly used to measure inter-rater reliability. 
This is useful to determine the risk of bias on the reliability of the classification.
Specifically, we created four statistically significant samples corresponding to the four independent researchers with a confidence level of 95 percent and a confidence interval of 5 percent.
Because each independent researcher classified on average 5795 issue sections, the size of each statistically significant sample is calculated to be 360 issue sections.
Then the first author analyzed the statistically significant samples independently and the Cohen's kappa coefficient between the independent researchers and the first author was calculated.
If the Cohen's kappa coefficient was below 0.7, the independent researcher discussed with the first author about discrepancies and subsequently reanalyzed all the issue sections in his/her subset.
Then the first author analyzed another statistically significant sample and Cohen's kappa was calculated again.
This process was repeated until the Cohen's kappa was above 0.7.

\subsection{Training and Executing Machine Learning Models}

\subsubsection{Machine Learning Models}

In order to accurately identify SATD in issues, we implement several machine learning approaches and compare their ability in identifying technical debt in issues.
We describe the approaches and why we selected them below:

\begin{itemize}
    \item \textbf{Traditional machine learning approaches (SVM, NBM, kNN, LR, RF)}:
    Support Vector Machine (SVM) \citep{sun2009strategies}, Naive Bayes Multinomial (NBM) \citep{mccallum1998comparison}, k-Nearest Neighbor (kNN) \citep{tan2006effective}, Logistic Regression (LR) \citep{genkin2007large}, and Random Forest (RF) \citep{xu2012improved} classifiers are widely used in text classification tasks \citep{kowsari2019text} due to their good classification accuracy. 
    Moreover, the results of current studies on SATD identification \citep{da2017using,huang2018identifying,flisar2019identification} show that these approaches achieve good accuracy in classifying SATD in source code comments.
    Thus, these approaches have potential to also achieve good accuracy when classifying SATD in issue trackers.
    Therefore, we train all of these classifiers using the implementation in Sklearn\footnote{\label{l:sklearn}\url{https://scikit-learn.org}} using Bag-of-Words (BoW) with default settings and compare their predictive performance.
    
    \item \textbf{Text Graph Convolutional Network (Text GCN)}:
    Text GCN generates a single large graph from the corpus and classifies text through classifying graph nodes using a graph neural network \citep{yao2019graph}.
    This approach achieves promising performance in text classification tasks by outperforming numerous state-of-the-art methods \citep{yao2019graph}.

    \begin{figure}[t]
      \centering
      \includegraphics[width=\columnwidth]{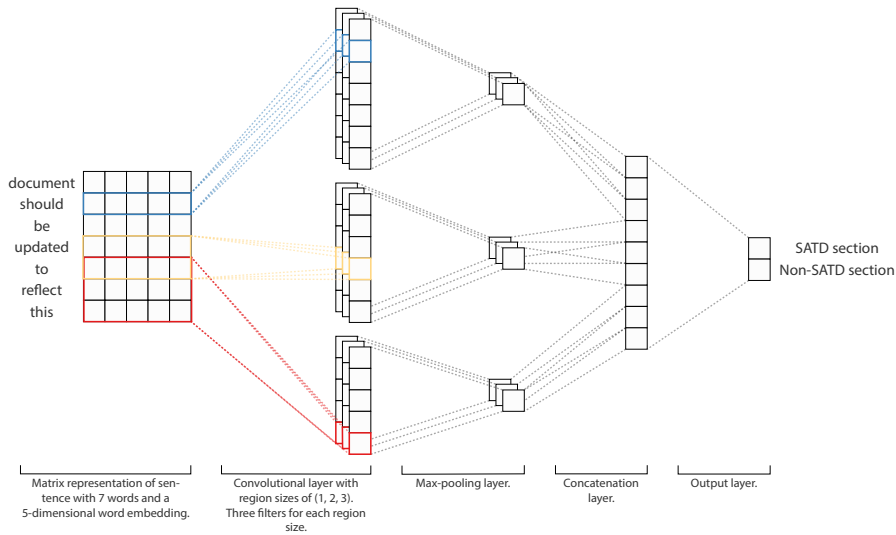}
      \caption{Architecture of the CNN model.}
      \label{f:cnn}
    \end{figure}
    
    \item \textbf{Text Convolutional Neural Network (Text CNN)}:
    Text CNN is a simple one-layer CNN proposed by \cite{kim2014convolutional}, that achieved high accuracy over the state of the art.
    The details of this approach are presented in more detail, as they are background knowledge for understanding some of the results in \cref{sec:results}. 
    The architecture of the model is presented in \cref{f:cnn}.
    The input issue section is first tokenized and converted into a matrix using an n-dimensional word embedding (see \cref{sec:wordembedding}).
    For example, in \cref{f:cnn}, the input issue section is \textit{`document should be updated to reflect this'}, which is represented as a $7 \times 5$ matrix because the issue section contains 7 words and the dimensionality of the word embedding is 5.
    Then the matrix is regarded as an image, and convolution operation is performed to extract the high level features.
    Because each row in the issue section matrix represents a word and the depth of the filter must be the same as the depth (width) of the input matrix, only the height of the filter can be adjusted, which is denoted by region size.
    It is important to note that multiple filters with different region sizes are applied to the issue section matrix to extract multiple features.
    In Fig. \ref{f:cnn}, the model adopts three filter region sizes (i.e., 1, 2, and 3) and three filters per region size. 
    Applying the three filter region sizes 1, 2, and 3 on the input issue section produces nine feature maps with the sizes of 7, 6, and 5. 
    For example, with a filter region size of 1, the convolution operation needs to be applied on every row (i.e. every word) of the input issue section, and thus producing a feature map with size of 7.
    After that, to make use of the information from each feature map, 1-max-pooling (which computes the maximum value of each feature map) is applied to extract a scalar from each feature.
    Then the output features are concatenated and flattened to form the penultimate layer.
    Finally, the output layer calculates the probability of the section to be a SATD section using the softmax activation function.
    This approach has been proven to be accurate for identifying SATD in source code comments \citep{ren2019neural}; thus it also has potential for accurately identifying SATD in issues. 
\end{itemize}

In this study, we mainly use Keras\footnote{\url{https://keras.io}} and Sklearn libraries\cref{l:sklearn} to implement machine learning approaches.
The machine learning models are trained on the NVIDIA Tesla V100 GPU.

\subsubsection{Baseline Approaches}

To compare the accuracy between the different classification approaches, we implement two baseline approaches.

\begin{itemize}
    \item \textbf{Baseline 1 (Random)}:
    This is a simple baseline approach, which assumes that the SATD detection is random.
    This random approach classifies sections as SATD sections randomly based on the probability of a section being a SATD section.
    For example, if 3,277 out of 23,180 sections are SATD sections in the training set, we assume the probability of a section being a SATD section is 14.1\%.
    Then the random approach randomly classifies any section in the test set as SATD section corresponding to the calculated probability (14.1\%).
    
    \item \textbf{Baseline 2 (Keyword)}:
    In the work of \cite{potdar2014exploratory}, they identified and summarized 62 SATD keywords, such as \textit{fixme}, \textit{ugly}, \textit{temporary solution}, \textit{this isn't quite right}, and \textit{this can be a mess}.
    Those keywords were used for automatically identifying SATD comments \citep{bavota2016large}.
    The SATD keyword-based method classifies a section as a SATD section when the section contains one or more of these SATD keywords.
\end{itemize}

\subsubsection{Strategies for Handling Imbalanced Data}
\label{sec:imbalanced_data}

As we can see in \cref{tb:projects}, about 15\% of the sections were classified as SATD sections in our issue dataset on average, which indicates that our dataset is imbalanced.
Imbalanced data always hinders the accuracy of the classifiers as the minority class tends to be overlooked \citep{fernandez2018learning}.
In order to improve the accuracy of machine learning models, we select the following three strategies for handing imbalanced data.

\begin{itemize}
    \item \textbf{Easy Data Augmentation (EDA)}:
    this technique augments text data through synonym replacement, random insertion, random swap, and random deletion \citep{wei2019eda}.
    To balance the dataset, we generate and add synthetic SATD sections to the training data using the EDA technique.
    
    \item \textbf{Oversampling}:
    This method simply replicates the minority class to re-balance the training data.
    We replicate the SATD sections to balance the SATD sections and non-SATD sections before training.
    
    \item \textbf{Weighted loss}:
    This method first calculates weights for all classes according to their occurrence.
    High frequency in occurrence leads to low weight value.
    Then the loss of each measurable element is scaled by the corresponding weight value in accordance with the class. 
    Weighted loss penalizes harder the wrongly classified sections from minority classes (i.e. false negative and false positive errors) during training of machine learning models to resolve the imbalanced data.
    This strategy is widely used for training CNN models on imbalanced datasets \citep{phan2017dnn,ren2019neural}.
\end{itemize}

\subsubsection{Word Embedding}
\label{sec:wordembedding}

Word embeddings refer to using a set of techniques mapping words to vectors of real numbers, which has been shown to boost the performance of text classification \citep{joulin2017bag,wieting2015towards}.
We choose word embedding technique for word representation since it is able to learn word meaning and semantics.
In this study, we train CNN models on top of four different word embeddings.
The first word embedding is simply initialized randomly.
The second word embedding is pre-trained by \cite{mikolov2018advances} on Wikipedia and news dataset.
The third word embedding is pre-trained by \cite{efstathiou2018word} on Stack Overflow posts, which is specific to the software engineering domain.
The last word embedding is trained by us on our collected issue data (summaries, descriptions, and comments) using the fastText technique \citep{mikolov2018advances} with default settings.

\subsubsection{Evaluation Metrics}

We use four statistics evaluating the accuracy of different approaches: true positive (TP) represents the number of sections correctly classified as SATD sections; false positive (FP) represents the number of sections classified as SATD sections when they are not SATD sections; true negative (TN) represents the number of sections correctly classified as not SATD sections; false negative (FN) represents the number of sections classified as not SATD sections when they are SATD sections.
Consequently, we calculate \textbf{precision} ($precision = \frac{TP}{TP+FP}$), \textbf{recall} ($recall = \frac{TP}{TP+FN}$), and \textbf{F1-score} ($F1 = 2 \times \frac{precision \times recall}{precision + recall}$).

The higher evaluation metric (i.e., precision, recall, or F1-score) means the better accuracy, whereas there is a trade-off between precision and recall.
In general, F1-score gives us an overall accuracy combining precision and recall.

\subsubsection{Keyword Extraction}

To better understand how developers declare technical debt in issues (to answer RQ2), we extract SATD keywords using the approach proposed by \cite{ren2019neural}.
This approach is built upon the CNN approach and is able to extract n-gram keywords using the backtracking technique.
More specifically, after feeding a SATD section to the CNN model, the most important features are selected according to their weights.
Then the corresponding filters are located via backtracking the selected features.
Finally, the n-gram keywords in the fed SATD section are located based on the filter position information.
In this study, we train the model on all seven projects' issue sections.
Then we summarize unigram to five-gram SATD keywords based on the extracted keywords.

\section{Results}
\label{sec:results}

\subsection{\textit{(RQ1.1) Which Algorithms Have the Best Accuracy to Capture Self-Admitted Technical Debt in Issue Tracking Systems?}}
\label{rq:1.1}

To evaluate the machine learning algorithms, we first combine all the issue sections from different projects, then shuffle and split the combined dataset into ten equally-sized partitions, while keeping the number of SATD sections approximately equal in all the partitions.
We then select one of the ten subsets for testing and the rest of the nine subsets for training, and repeat this process for each subset and calculate the average precision, recall, and F1-score over ten experiments.
This is called stratified 10-fold cross-validation.
\cref{tb:f1_all_methods} shows the precision, recall, and F1-score of deep learning approaches (Text CNN and Text GCN), traditional machine learning approaches (SVM, NBM, kNN, LR, and RF), and baseline approaches (keyword-based and random).
The best results are highlighted in bold.

\begin{table}[t]
\caption{Comparison of precision, recall, and F1-score between machine learning and baseline approaches.}
\label{tb:f1_all_methods}
\begin{center}
\resizebox{\columnwidth}{!}{
\def\arraystretch{1.2}
\begin{tabular}{@{\extracolsep{4pt}}C{1.5cm}C{2.65cm}C{1.4cm}C{1.1cm}C{1.3cm}C{1.6cm}C{1.6cm}@{}@{}}
\hline
\textbf{Type} & \textbf{Classifier} & \textbf{Precision} & \textbf{Recall} & \textbf{F1-score} & \textbf{F1-score Imp. Over Random} & \textbf{F1-score Imp. Over Keyword} \\
\hline
\multirow{4}{*}{\makecell{Deep\\Learning}} & \textbf{Text CNN (rand)} & 0.685 & 0.530 & \textbf{0.597} & \textbf{4.3$\times$} & \textbf{13.6$\times$} \\
 & \textbf{Text CNN (wiki)} & 0.677 & 0.463 & 0.549 & 3.9$\times$ & 12.5$\times$ \\
 & \textbf{Text CNN (SO)} & 0.651 & \textbf{0.541} & 0.590 & 4.2$\times$ & 13.4$\times$ \\
 & \textbf{Text GCN} & 0.474 & 0.056 & 0.081 & 0.6$\times$ & 1.8$\times$ \\
\hline
\multirow{5}{*}{\makecell{Traditional\\Machine\\Learning}} & \textbf{SVM} & \textbf{0.861} & 0.179 & 0.295 & 2.1$\times$ & 6.7$\times$ \\
 & \textbf{NBM} & 0.520 & 0.539 & 0.529 & 3.8$\times$ & 12.0$\times$ \\
 & \textbf{kNN} & 0.582 & 0.029 & 0.055 & 0.4$\times$ & 1.2$\times$ \\
 & \textbf{LR} & 0.643 & 0.430 & 0.515 & 3.7$\times$ & 11.7$\times$ \\
 & \textbf{RF} & 0.730 & 0.182 & 0.291 & 2.1$\times$ & 6.6$\times$ \\
\hline
\multirow{2}{*}{\makecell{Baseline}} & \textbf{Random} & 0.140 & 0.139 & 0.139 & & \\
 & \textbf{Keyword} & 0.515 & 0.023 & 0.044 & & \\
\hline
\end{tabular}
}
\end{center}
\end{table}

As we can see in \cref{tb:f1_all_methods}, Text CNN with randomized word embeddings achieves the highest average F1-score of 0.597.
This contrasts earlier evidence \citep{kim2014convolutional}, where Text CNN with two pre-trained word embeddings (Wiki-news and StackOverflow-post word embeddings) degraded the model's predictive accuracy.
It is also important to note that, Text CNN with the word embeddings trained specifically for the software engineering domain (i.e., the StackOverflow-post word embeddings) is still slightly worse than the random word embeddings on the average F1-score (0.590 and 0.597 respectively).
Among the traditional machine learning techniques, NBM and LR achieve decent F1-scores of 0.529 and 0.515.

Observing \cref{tb:f1_all_methods}, we also notice that most of the approaches achieve decent precisions. 
While SVM and RF achieve the two highest average precisions of 0.861 and 0.730, their average recalls are relatively poor (0.295 and 0.291 respectively).
Moreover, we notice that the keyword-based method achieves an average precision of 0.515; this means that the SATD sections in source code comments and issue tracking systems share some similarities and the keywords from source code comments (see \cite{potdar2014exploratory} indicating SATD, such as \textit{fixme}, \textit{ugly}, and \textit{temporary solution}) are useful for detecting SATD sections in issues.
However, the recall of the keyword-based method is the lowest (0.023) among all methods, which might result from the low coverage of these source code comments keywords.

\begin{framed}
\noindent \textit{Text CNN with default settings using random word embeddings achieves the highest F1-score (0.597) among all approaches, followed by Text CNN with default settings using pre-trained StackOverflow-post word embeddings (0.590).
}
\end{framed}

\subsection{\textit{(RQ1.2)} How to Improve Accuracy of Machine Learning Model?}
\label{sec:rq1.2}

After establishing that the Text CNN approach achieves the highest F1-score among all approaches, we can further investigate its improvement. 
In the following sub-sections, we investigate handling imbalanced data, refining word embeddings and tuning CNN hyperparameters.

\begin{table}[b]
\caption{Comparison of average precision, recall, and F1-score between different imbalanced data handling strategies.}
\label{tb:imbalance_data}
\begin{center}
\resizebox{\columnwidth}{!}{
\def\arraystretch{1.2}
\begin{tabular}{C{2.1cm}C{2.6cm}C{1.5cm}C{1.5cm}C{1.5cm}C{1.5cm}}
\hline
\textbf{Method} & \textbf{Word Embedding} & \textbf{Precision (Average)} & \textbf{Recall (Average)} & \textbf{F1-score (Average)} & \textbf{F1-score Imp.} \\
\hline
\multirow{4}{*}{\textbf{Default}} & Random & 0.685 & 0.530 & 0.597 & - \\
 & Wiki-news & 0.677 & 0.463 & 0.549 & - \\
 & StackOverflow-post & 0.651 & 0.541 & 0.590 & - \\
\cline{2-6}
 & Average & \textbf{0.671} & 0.511 & 0.578 & - \\
\hline
\multirow{4}{*}{\textbf{EDA}} & Random & 0.606 & 0.470 & 0.529 & -11.3\% \\
 & Wiki-news & 0.556 & 0.406 & 0.469 & -14.5\% \\
 & StackOverflow-post & 0.604 & 0.595 & 0.599 & 1.5\% \\
\cline{2-6}
 & Average & 0.588 & 0.490 & 0.532 & -7.9\% \\
\hline
\multirow{4}{*}{\textbf{Oversampling}} & Random & 0.573 & 0.717 & 0.636 & 6.5\% \\
 & Wiki-news & 0.591 & 0.592 & 0.591 & 7.6\% \\
 & StackOverflow-post & 0.610 & 0.618 & 0.612 & 3.7\% \\
\cline{2-6}
 & Average & 0.591 & 0.642 & 0.613 & 6.0\% \\
\hline
\multirow{4}{*}{\textbf{Weighted loss}} & Random & 0.555 & 0.735 & 0.632 & 5.8\% \\
 & Wiki-news & 0.583 & 0.617 & 0.599 & 9.1\% \\
 & StackOverflow-post & 0.591 & 0.640 & 0.613 & 3.8\% \\
\cline{2-6}
 & Average & 0.576 & \textbf{0.664} & \textbf{0.614} & \textbf{6.2\%} \\
\hline
\end{tabular}
}
\end{center}
\end{table}

\subsubsection{Handling Imbalanced Data}

As we can see in \cref{tb:projects}, on average only 14.1\% sections were classified as SATD sections in our issue dataset, which indicates that our dataset is imbalanced.
Since the Text CNN approach is not designed for imbalanced data classification tasks, firstly we look into methods for handling imbalanced data.
We train the Text CNN model with different word embeddings using the aforementioned imbalanced data handling techniques discussed in \cref{sec:imbalanced_data}.

\cref{tb:imbalance_data} presents the precision, recall, and F1-score improvement of Text CNN approach with different imbalanced data handling techniques. 
We can see that EDA always has a negative effect on training and degrades the F1-score by 7.9\% on average.
The other two imbalanced data handling techniques (i.e., oversampling and weighted loss) achieve a similar improvement on three different word embeddings, while the average F1-score improvement is 6.0\% and 6.2\% respectively. 
Since oversampling replicates the sections in minority classes, the training dataset using oversampling is significantly larger than the one using weighted loss.
Thus, we notice that the time spent on training using oversampling is 30.8\% longer than using weighted loss.
Therefore, we choose weighted loss as the imbalanced data handling technique.

\begin{table}[t]
\caption{Comparison of average precision, recall, and F1-score between different word vector settings using weighted loss.}
\label{tb:word_vectors}
\begin{center}
\resizebox{\columnwidth}{!}{
\def\arraystretch{1.2}
\begin{tabular}{C{2.8cm}C{2.5cm}C{1.5cm}C{1.5cm}C{1.5cm}}
\hline
\textbf{Word Embedding} & \textbf{Dimensionality} & \textbf{Precision (Average)} & \textbf{Recall (Average)} & \textbf{F1-score (Average)} \\
\hline
Random & 300 & 0.555 & \textbf{0.735} & 0.632 \\
Wiki-news & 300 & 0.583 & 0.617 & 0.599 \\
StackOverflow-post & 200 & 0.591 & 0.640 & 0.613 \\
\hline
 & 100 & 0.647 & 0.686 & 0.664 \\
Issue-tracker-data & 200 & \textbf{0.662} & 0.680 & 0.670 \\
 & 300 & 0.648 & 0.703 & \textbf{0.673} \\
\hline
\end{tabular}
}
\end{center}
\end{table}

\subsubsection{Refining Word Embeddings}

In \cref{tb:imbalance_data}, we notice that the average F1-score is not improved using pre-trained word embeddings (i.e., Wiki-news and StackOverflow-post) compared to using Random word embedding.
Thus, we train the word embeddings on our issue dataset using fastText technique \citep{mikolov2018advances} with the dimension size of word embeddings setting to 100, 200 and 300.
Because fastText is able to learn subword information during word representation training \citep{bojanowski2017enriching}, we use it to train word embeddings.
As can be seen in \cref{tb:word_vectors}, we observe that the word embeddings trained on our dataset with different dimension sizes all outperform the Random word embeddings and pre-trained word embeddings (i.e., Wiki-news and StackOverflow-post) significantly.
Besides, while increasing the size of dimensions, the F1-score slightly grows.
Therefore, we use the word embeddings trained on our dataset and choose 300 as the dimension size.

\begin{table}[t]
\caption{Comparison of average precision, recall, and F1-score between different filter region size settings using issue-tracker-data word embeddings and weighted loss.}
\label{tb:region_size}
\begin{center}
\resizebox{0.9\columnwidth}{!}{
\def\arraystretch{1.2}
\begin{tabular}{C{1.5cm}C{3.5cm}C{1.5cm}C{1.5cm}C{1.5cm}}
\hline
\textbf{Type} & \textbf{Region Size} & \textbf{Precision (Average)} & \textbf{Recall (Average)} & \textbf{F1-score (Average)} \\
\hline
\multirow{4}{*}{\textbf{Single}} & \textbf{(1)} & 0.552 & \textbf{0.782} & 0.646 \\
 & \textbf{(3)} & 0.638 & 0.707 & \textbf{0.670} \\
 & \textbf{(5)} & 0.631 & 0.686 & 0.657 \\
 & \textbf{(7)} & \textbf{0.642} & 0.665 & 0.652 \\
\hline
\multirow{12}{*}{\textbf{Multiple}} & \textbf{(1,2)} & 0.643 & \textbf{0.715} & 0.676 \\
 & \textbf{(1,2,3)} & 0.657 & 0.711 & \textbf{0.682} \\
 & \textbf{(2,3,4)} & 0.655 & 0.706 & 0.677 \\
 & \textbf{(3,4,5)} & 0.648 & 0.703 & 0.673 \\
 & \textbf{(1,2,3,4)} & 0.663 & 0.703 & 0.679 \\
 & \textbf{(1,3,5,7)} & 0.675 & 0.677 & 0.675 \\
 & \textbf{(2,4,6,8)} & 0.674 & 0.665 & 0.669 \\
 & \textbf{(1,2,3,4,5)} & 0.678 & 0.685 & 0.680 \\
 & \textbf{(1,2,3,5,7)} & \textbf{0.681} & 0.682 & 0.680 \\
 & \textbf{(1,3,4,5,7)} & 0.667 & 0.686 & 0.676 \\
 & \textbf{(1,3,5,7,9)} & 0.668 & 0.681 & 0.673 \\
 & \textbf{(1,2,3,4,5,6)} & 0.669 & 0.691 & 0.678 \\
 & \textbf{(1,2,3,4,5,6,7)} & 0.669 & 0.680 & 0.673 \\
\hline
\end{tabular}
}
\end{center}
\end{table}

\subsubsection{Tuning CNN Hyperparameters}

We follow the guidelines provided by \cite{zhang2017sensitivity} to tune the hyperparameters of the CNN model.
We first conduct a line-search over the single filter region size (i.e., setting the region size to (1), (3), (5), and (7)) to find the single filter region size with the best accuracy.
The results are reported in \cref{tb:region_size}.
As we can see, in this study, the single region size (3) outperforms other single region sizes.

Second, we further explore the effect of multiple region sizes using regions sizes near this single best size (3) according to the suggestion from the guidelines; this is because combing multiple filters using region sizes near the best size always results in better accuracy compared to only using single best region size \citep{zhang2017sensitivity}.
Because we cannot enumerate all combinations of region sizes, we consider the combinations of region sizes of (1,2), (1,2,3), (2,3,4), (3,4,5), (1,2,3,4), (1,3,5,7), (2,4,6,8), (1,2,3,4,5), (1,2,3,5,7), (1,3,4,5,7), (1,3,5,7,9), (1,2,3,4,5,6), and (1,2,3,4,5,6,7).
From the results in \cref{tb:region_size}, we can see that among all combinations, (1,2,3) achieves the highest F1-score, followed by (1,2,3,4,5) and (1,2,3,5,7) by small margins (0.682, 0.680 and 0.680 respectively).
Thus, we choose (1,2,3) as the region size.

\begin{table}[b]
\caption{Comparison of average precision, recall, and F1-score between different filter region size settings using issue-tracker-data word embeddings and weighted loss and setting multiple region size to (1, 2, 3).}
\label{tb:num_feature}
\begin{center}
\resizebox{0.8\columnwidth}{!}{
\def\arraystretch{1.2}
\begin{tabular}{C{4cm}C{1.5cm}C{1.5cm}C{1.5cm}}
\hline
\textbf{Number of Feature Maps} & \textbf{Precision (Average)} & \textbf{Recall (Average)} & \textbf{F1-score (Average)} \\
\hline
\textbf{50} & 0.640 & \textbf{0.713} & 0.674 \\
\textbf{100} & 0.657 & 0.711 & 0.682 \\
\textbf{200} & \textbf{0.685} & 0.689 & \textbf{0.686} \\
\textbf{400} & 0.675 & 0.699 & 0.685 \\
\textbf{600} & 0.677 & 0.671 & 0.683 \\
\hline
\end{tabular}
}
\end{center}
\end{table}

Third, we investigate the effect of the number of feature maps for filter region size.
We keep other settings constant and only change the number of feature maps relative to the default number of features 100.
We set the number of features to 50, 100, 200, 400, and 600 under the aforementioned guidelines \citep{zhang2017sensitivity}.
The result is demonstrated in \cref{tb:num_feature}.
As we can see, while increasing the number of feature maps, the average F1-score is slightly improved until 200, so we set 200 as the number of feature maps.

\subsubsection{Final Results After Machine Learning Optimization}

To conclude our optimization, after handling imbalanced data, refining word embeddings, and tuning hyperparameters, we compare the precision, recall, and F1-score of our customized CNN approach against the Text CNN with default settings.
The results indicate that our customized approach improves the recall and F1-score over the Text CNN with the default settings by 30.0\% and 14.9\% respectively.
More specifically, on average it does not improve precision (0.685); the recall of our approach (0.689) increases by 30.0\% compared to the original approach (0.530).
Overall, our customized CNN approach improves the F1-score by 14.9\% from 0.597 to 0.686 on average. 

\begin{framed}
\noindent \textit{After handling imbalanced data, refining word embeddings, and tuning hyperparameters, our customized CNN approach improves the F1-score over the Text CNN with default settings by 14.9\% from 0.597 to 0.686.
}
\end{framed}

\subsection{\textit{(RQ1.3) How Can Transfer Learning Improve the Accuracy of Identifying Self-Admitted Technical Debt in Issue Tracking Systems?}}
\label{sec:rq1.3}

To leverage the knowledge in existing SATD datasets, we follow the transfer learning guidelines on text classification provided by \cite{semwal2018practitioners}.
Specifically, Semwal \textit{et al.} conducted a series of experiments and presented the cases and settings that could lead to a positive transfer (i.e. the transfer results in improved accuracy).

In this study, we explore how the accuracy of the CNN model for SATD identification can be further improved through transfer learning.
As can be seen in \cref{f:cnn}, The CNN model consists of an embedding layer, a convolutional layer, a max-pooling layer, a concatenation layer, and an output layer.
The embedding layer is responsible for converting words to n-dimensional vectors, which is pre-trained on various datasets in this work.
The convolutional (C) layer plays a critical role in learning various kinds of features.
The weights of its kernels are trainable and transferable.
The max-pooling layer is a regular layer that down-samples the input representation.
The concatenation layer is used for concatenating the inputs along a specified dimension.
The output (O) layer is in charge of producing the final result.
The weight and biases in the output layer are trainable and transferable.
Because there are no trainable parameters (i.e. the parameters are fixed during training) in the embedding layer, max-pooling layer, and concatenation layer, we explore the transfer learning settings for the convolutional layer (C) and output layer (O).

\begin{table}[t]
\caption{Comparison of average precision, recall, and F1-score between different transfer learning settings (C\faUnlock~O\faUnlock~ and C\faUnlock~O\faAsterisk) and without transfer learning setting (C\faAsterisk~O\faAsterisk).}
\label{tb:transfer_learning_other}
\begin{center}
\resizebox{0.9\columnwidth}{!}{
\def\arraystretch{1.2}
\begin{tabular}{C{2.5cm}C{2cm}C{1.5cm}C{1.5cm}C{1.5cm}}
\hline
\textbf{Source Dataset} & \textbf{Setting} & \textbf{Precision} & \textbf{Recall} & \textbf{F1-score} \\
\hline
- & C\faAsterisk~O\faAsterisk & 0.685 & 0.689 & 0.686 \\
\hline
\multirow{2}{*}{CO-SATD}  & C\faUnlock~O\faUnlock & 0.674 & 0.671 & 0.672 \\ 
 & C\faUnlock~O\faAsterisk & 0.675 & 0.684 & 0.679 \\
\hline
\multirow{2}{*}{AMZ2}  & C\faUnlock~O\faUnlock & 0.666 & 0.664 & 0.665 \\ 
 & C\faUnlock~O\faAsterisk & 0.681 & 0.664 & 0.672 \\
\hline
\multirow{2}{*}{YELP2}  & C\faUnlock~O\faUnlock & 0.670 & 0.677 & 0.673 \\ 
 & C\faUnlock~O\faAsterisk & \textbf{0.689} & 0.677 & 0.681 \\
\hline
\multirow{2}{*}{JIRA-SEN}  & C\faUnlock~O\faUnlock & 0.676 & \textbf{0.696} & 0.684 \\ 
 & C\faUnlock~O\faAsterisk & \textbf{0.689} & 0.694 & \textbf{0.691} \\
\hline
\multirow{2}{*}{SO-SEN}  & C\faUnlock~O\faUnlock & 0.682 & 0.686 & 0.683 \\ 
 & C\faUnlock~O\faAsterisk & 0.685 & 0.685 & 0.685 \\
\hline
\end{tabular}
}
\end{center}
\end{table}

To transfer the knowledge from the source domain to aid the target domain, we initialize the parameters of the target model to be trained with the parameters trained on the transfer learning source dataset.
To avoid confusion, we use the same nomenclature as the work by \cite{semwal2018practitioners}.
There are three settings for parameters during transfer learning:

\begin{itemize}
    \item \faAsterisk: Parameters are not transferred, but are randomly initialized and allowed to fine-tune.
    \item \faUnlock: Parameters are transferred and allowed to fine-tune during training.
    \item \faLock: Parameters are transferred and frozen, i.e., they can not learn during training.
\end{itemize}

According to the transfer learning guideline \citep{semwal2018practitioners}, we use the fine-tuning (\faUnlock) setting for the convolutional (C) layer, and either transferred parameters for fine-tuning (\faUnlock) or random initialization for fine-tuning (\faAsterisk) setting for the output (O) layer, which results in two combinations of settings (i.e., C\faUnlock~O\faUnlock~ and C\faUnlock~O\faAsterisk).

Thus, we conduct experiments with the aforementioned two transfer learning settings on our issue dataset: a) transferring and allowing both the convolutional layer and output layer to fine-tune (C\faUnlock~O\faUnlock); b) only transferring and allowing the fine-tuning of the convolutional layer and randomizing the parameters in the output layer for fine-tuning (C\faUnlock~O\faAsterisk).
Besides, we compare the results with learning from scratch (C\faAsterisk~O\faAsterisk).
For the source dataset selection, we choose five datasets as the transfer learning source datasets: 

\begin{itemize}
    \item \textit{Source code comment SATD dataset (CO-SATD)} \citep{da2017using}: 
    We chose this dataset, because it contains SATD in source code comments, which is highly similar to our issue SATD dataset.
    This dataset contains 62,566 comments, in which 4,071 comments were annotated as SATD comments.
    
    \item \textit{Amazon review (AMZ2)} and \textit{Yelp review (YELP2) datasets} \citep{zhang2015character}:
    AMZ2 contains 2,000,000 Amazon product reviews for each polarity.
    YELP2 includes 289,900 business reviews for each polarity. 
    We selected these two datasets, because their size is significantly bigger than our dataset and because they are commonly used for text classification tasks \citep{semwal2018practitioners}.
    
    \item \textit{Jira issue sentiment (JIRA-SEN)} and \textit{Stack Overflow post sentiment (SO-SEN) datasets} \citep{ortu2016emotional,calefato2018sentiment}: 
    Both datasets are relatively small datasets (only containing 926 and 2,728 samples respectively). 
    We chose these two datasets because they are in the software engineering domain.
\end{itemize}

To evaluate the efficiency of transfer learning, we still use stratified 10-fold cross-validation.
We first train our model on these transfer learning source datasets individually and then retrain the models on our issue SATD dataset in \cref{tb:transfer_learning_other} with the transfer learning settings (C\faUnlock~O\faUnlock~ and C\faUnlock~O\faAsterisk).
The results are illustrated in \cref{tb:transfer_learning_other}.
We observe that applying transfer learning with JIRA-SEN and SO-SEN outperform CO-SATD, AMAZ2, and YELP2 with respect to F1-score. 
This indicates that sentiment information in software engineering domain could be useful for SATD identification.
Moreover, only the F1-score achieved by JIRA-SEN using the setting (C\faUnlock~O\faAsterisk) outperforms training from scratch setting (C\faAsterisk~O\faAsterisk), which indicates positive transfers \citep{perkins1992transfer}.
The F1-score is slightly improved from 0.686 to 0.691 by leveraging the JIRA-SEN dataset.
Our findings here show that transfer learning could improve the F1-score of SATD identification in issue tracking systems, but it highly depends on the source dataset selection.

\begin{table}[b]
\caption{F1-score of different transfer learning settings with different number of training sections from 0 to 900. (\faAsterisk: randomly initialize and allow to fine-tune parameters; \faUnlock: allow to fine-tune transferred parameters.)}
\label{tb:transfer_learning}
\begin{center}
\resizebox{\columnwidth}{!}{
\def\arraystretch{1.2}
\begin{tabular}{@{\extracolsep{4pt}}C{1.4cm}C{1.1cm}C{0.79cm}C{0.79cm}C{0.79cm}C{0.79cm}C{0.79cm}C{0.79cm}C{0.79cm}C{0.79cm}C{0.79cm}C{0.79cm}@{}@{}}
\hline
\multirow{2}{*}{\textbf{\makecell{Source \\Dataset}}} & \multirow{2}{*}{\textbf{Setting}} & \multicolumn{10}{c}{\textbf{Number of Issue Sections Used for Training}} \\
\cline{3-12}
 &  & 0 & 100 & 200 & 300 & 400 & 500 & 600 & 700 & 800 & 900 \\
\hline
- & C\faAsterisk~O\faAsterisk & 0.157 & 0.357 & 0.386 & 0.424 & 0.451 & 0.474 & 0.488 & 0.501 & 0.507 & 0.523 \\
\hline
\multirow{2}{*}{CO-SATD} & C\faUnlock~O\faUnlock & 0.260 & \textbf{0.395} & \textbf{0.402} & 0.412 & 0.425 & 0.435 & 0.442 & 0.445 & 0.458 & 0.461 \\
 & C\faUnlock~O\faAsterisk & 0.202 & 0.367 & 0.393 & \textbf{0.425} & \textbf{0.469} & \textbf{0.478} & \textbf{0.493} & 0.497 & \textbf{0.518} & 0.515 \\
\hline
\multirow{2}{*}{AMZ2} & C\faUnlock~O\faUnlock & 0.170 & 0.289 & 0.295 & 0.305 & 0.334 & 0.345 & 0.361 & 0.364 & 0.380 & 0.388 \\
 & C\faUnlock~O\faAsterisk & 0.145 & 0.349 & 0.386 & 0.400 & 0.438 & 0.451 & 0.461 & 0.475 & 0.487 & 0.497 \\
\hline
\multirow{2}{*}{YELP2} & C\faUnlock~O\faUnlock & 0.051 & 0.288 & 0.290 & 0.300 & 0.323 & 0.329 & 0.352 & 0.358 & 0.378 & 0.389 \\
 & C\faUnlock~O\faAsterisk & 0.147 & 0.341 & 0.371 & 0.398 & 0.439 & 0.462 & 0.477 & 0.481 & 0.506 & 0.500 \\
\hline
\multirow{2}{*}{JIRA-SEN} & C\faUnlock~O\faUnlock & \textbf{0.273} & 0.337 & 0.352 & 0.379 & 0.411 & 0.419 & 0.435 & 0.449 & 0.468 & 0.473 \\
 & C\faUnlock~O\faAsterisk & 0.169 & 0.354 & 0.391 & 0.410 & 0.452 & 0.472 & 0.486 & \textbf{0.503} & 0.517 & \textbf{0.531} \\
\hline
\multirow{2}{*}{SO-SEN} & C\faUnlock~O\faUnlock & 0.256 & 0.296 & 0.301 & 0.330 & 0.356 & 0.371 & 0.384 & 0.393 & 0.418 & 0.428 \\
 & C\faUnlock~O\faAsterisk & 0.089 & 0.350 & 0.389 & 0.410 & 0.458 & 0.477 & 0.484 & 0.498 & 0.515 & 0.517 \\
\hline
\end{tabular}
}
\end{center}
\end{table}

Furthermore, we investigate the effectiveness of transfer learning when data is insufficient for training the target model.
We report the F1-score achieved by two transfer learning settings (C\faUnlock~O\faUnlock~ and C\faUnlock~O\faAsterisk) versus learning from scratch (C\faAsterisk~O\faAsterisk) on the training dataset containing 0 to 900 issue sections in \cref{tb:transfer_learning}.
From the results, we find that when training data is extremely scarce (less than 200 issue sections for training), transferring both the convolutional layer and output layer (C\faUnlock~O\faUnlock) performs best by selecting CO-SATD or JIRA-SEN as the transfer learning source dataset.
When there is sightly more training data (between 300 and 600 sections for training), only transferring the convolution layer (C\faUnlock~O\faAsterisk) achieves the highest F1-score by using CO-SATD as the transfer learning source dataset.
When more training data is available (more than 700 sections for training), the model without transferred parameters (C\faAsterisk~O\faAsterisk) likely starts to outperforms others by using CO-SATD or JIRA-SEN as the transfer learning source dataset.

\begin{framed}
\noindent \textit{Using Jira issue sentiment dataset (JIRA-SEN) as the source dataset for transfer learning, improves the F1-score but only slightly (from 0.686 to 0.691) with the setting of C\faUnlock~O\faAsterisk.
When training data is insufficient, using Jira issue sentiment dataset (JIRA-SEN) or source code comment SATD dataset (CO-SATD) as the source dataset could boost the F1-score.
}
\end{framed}

\subsection{\textit{(RQ2) Which Keywords Are the Most Informative to Identify Self-Admitted Technical Debt in Issue Tracking Systems?}}
\label{rq:2}

We first summarize a list of top SATD keywords (unigram to five-gram) based on the extracted keywords.
The results are shown in \cref{tb:n-grams}.
Note that keywords which are similar to the keywords in source code comments (see \cite{potdar2014exploratory}) are underlined.
The first author manually linked the keywords to types and indicators of SATD by checking the types and indicators of issue sections that contain the keywords based on the existing classification framework from our previous work \citep{9226330} (see \cref{tb:statistics_type_projects}).
Subsequently, the other authors checked and confirmed the correlation between keywords with types and indicators.
We can see that our keywords are intuitive and potentially indicate the types and indicators of the technical debt; the types are shown in \cref{tb:n-grams} and \cref{tb:keyword_satd_link} while the definitions of types and indicators are included in \cref{tb:statistics_type_projects}.

\begin{table}[t]
\caption{Extracted top n-gram keywords from uni- to five-gram.}
\label{tb:n-grams}
\begin{center}
\resizebox{\columnwidth}{!}{
\def\arraystretch{1.2}
\begin{tabular}{L{4cm}L{4cm}L{4cm}}
\hline
\textbf{Unigram Keyword} & \textbf{Bigram Keyword} & \textbf{Trigram Keyword} \\
\hline
flaky (test) & too much & \underline{get rid of} \\
leak (code) & not used (code) & not thread safe (requirement) \\
unused (code) & more readable (code) & \underline{clean up code} (code) \\
unnecessary (code) & more efficient (code) & not done yet (requirement) \\
typo (code/documentation) & dead code (code) & avoid extra seek \\
slow (code) & infinite loop (code) & reduce duplicate code (code) \\
redundant (code) & too long (documentation) & no longer needed (code) \\
confusing (code/documentation) & not implemented (requirement) & not supported yet (requirement) \\
nit & less verbose (code) & documentation doesn't match (documentation) \\
\underline{ugly} & more robust (design) & \underline{short term solution} \\
simplify (code/documentation) & speed up (code) & spurious error messages (code) \\
misleading (documentation) & missing documentation (documentation) & it'd be nice \\
\hline
\end{tabular}
}
\resizebox{\columnwidth}{!}{
\def\arraystretch{1.2}
\begin{tabular}{L{6cm}L{6cm}}
\textbf{Four-gram Keyword} & \textbf{Five-gram Keyword} \\
\hline
please add a test (test) & wastes a lot of space \\
would significantly improve performance (code) & there is no unit test (test) \\
makes it much easier & lead to huge memory allocation (design) \\
avoid calling it twice (code) & test doesn't add much value (test) \\
takes a long time (code) & some holes in the doc (documentation) \\
good to have coverage (test) & by hard coding instead of (code) \\
makes it very hard & should be updated to reflect (documentation) \\
patch doesn't apply cleanly (code) & more tightly coupled than ideal (design) \\
it's not perfectly documented (documentation) & any chance of a test (test) \\
need to update documentation (documentation) & should improve a bit by \\
make it less brittle (design) & it'd help code readability if (code) \\
documentation does not mention (documentation) & solution won't be really satisfactory \\
\hline
\end{tabular}
}
\end{center}
\end{table}

\begin{table}[thpb]
\caption{Correlation between keywords and types/indicators of SATD.}
\label{tb:keyword_satd_link}
\begin{center}
\resizebox{\columnwidth}{!}{
\def\arraystretch{1.2}
\begin{tabular}{@{\extracolsep{4pt}}L{0.5
cm}L{2.2cm}L{4.3cm}L{4.55cm}@{}@{}}
\hline
\multicolumn{2}{l}{\textbf{Type Indicator}} & \textbf{Keyword} & \textbf{Example} \\
\hline
\parbox[t]{2mm}{\multirow{23}{*}{\rotatebox[origin=c]{90}{\textbf{Code}}}} & \multirow{3}{*}{Complex Code} & simplify & \multirow{3}{4.5cm}{\textit{``That can \textbf{simplify} the logic there.''} - [HADOOP-10295]} \\
 &  & redundant &  \\
 &  & less verbose &  \\
 \cline{2-3}
 & \multirow{5}{*}{Dead Code} & unused & \multirow{5}{4.5cm}{\textit{``I would like to remove this as its \textbf{no longer needed}, and also its code is not complete.''} - [Camel-8174]} \\
 &  & unnecessary &  \\
 &  & not used &  \\
 &  & dead code &  \\
 &  & no longer needed &  \\
 \cline{2-3}
 & Duplicated Code & reduce duplicate code &  \\
 \cline{2-3}
 & \multirow{10}{*}{\makecell[l]{Low-Quality \\Code}} & typo & \multirow{8}{4.5cm}{\textit{``...to make their code \textbf{more readable}. I would like to see something like this in the API...''} - [HBase-1990]} \\
 &  & leak &  \\
 &  & confusing &  \\
 &  & more readable &  \\
 &  & infinite loop &  \\
 &  & clean up code &  \\
 &  & spurious error messages &  \\
 &  & avoid calling it twice &  \\
 &  & patch doesn't apply cleanly &  \\
 &  & it'd help code readability if &  \\
 &  & by hard coding instead of &  \\
 \cline{2-3}
 & \multirow{6}{*}{Slow Algorithm} & slow & \multirow{5}{4.5cm}{\textit{``Rowlocks should use rentHashMap as it is much \textbf{more efficient} than Collections.synchronizedMap(HashMap)''} - [HBase-798]} \\
 &  & more efficient &  \\
 &  & speed up &  \\
 &  & would significantly improve performance &  \\
 &  & takes a long time &  \\
\hline
\parbox[t]{2mm}{\multirow{4}{*}{\rotatebox[origin=c]{90}{\textbf{Design}}}} & \multirow{4}{*}{\makecell[l]{Non-Optimal \\Decision}} & more robust & \multirow{4}{4.5cm}{\textit{``...didn't tackle those pieces yet. They also seem \textbf{more tightly coupled than ideal}.''} - [HBase-12749]} \\
 &  & make it less brittle &  \\
 &  & lead to huge memory allocation &  \\
 &  & more tightly coupled than ideal &  \\
\hline
\parbox[t]{2mm}{\multirow{6}{*}{\rotatebox[origin=c]{90}{\textbf{Requirement}}}} & \multirow{3}{*}{\makecell[l]{Requirement \\Partially \\Implemented}} & not implemented & \multirow{3}{4.5cm}{\textit{``\textbf{Not implemented} reached in virtual void...''} - [Chromium-43196]} \\
 &  & not done yet &  \\
 &  & not supported yet &  \\
 \cline{2-3}
 & Non-Functional Requirements Not Satisfied & Not thread safe & \textit{``AuthorizationPolicy is \textbf{not thread-safe}''} - [Impala-7682]\\
\hline
\parbox[t]{2mm}{\multirow{12}{*}{\rotatebox[origin=c]{90}{\textbf{Documentation}}}} & \multirow{7}{*}{\makecell[l]{Outdated \\Documentation}} & missing documentation & \multirow{7}{4.5cm}{\textit{``I am using this opportunity to fill in \textbf{some holes in the doc}...''} - [Impala-991]} \\
 &  & documentation doesn't match &  \\
 &  & some holes in the doc &  \\
 &  & it's not perfectly documented &  \\
 &  & need to update documentation &  \\
 &  & documentation does not mention &  \\
 &  & should be updated to reflect &  \\
 \cline{2-3}
 & \multirow{5}{*}{\makecell[l]{Low-Quality \\Documentation}} & typo & \multirow{5}{4.5cm}{\textit{`Default searches documentation \textbf{misleading} about single-change search match behaviour in UI''} - [Gerrit-8592]} \\
 &  & confusing &  \\
 &  & simplify &  \\
 &  & misleading &  \\
 &  & too long &  \\
\hline
\parbox[t]{2mm}{\multirow{6}{*}{\rotatebox[origin=c]{90}{\textbf{Test}}}} & \multirow{3}{*}{Lack of Tests} & please add a test & \multirow{3}{4.5cm}{\textit{``It looks good to me except these. \textbf{Please add a test} case for the code change...''} - [Hadoop-12155]} \\
 &  & there is no unit test &  \\
 &  & any chance of a test &  \\
 \cline{2-3}
 & \multirow{2}{*}{Low Coverage} & good to have coverage & \multirow{3}{4.5cm}{\textit{``this \textbf{test doesn't add much value}, does it?''} - [Gerrit-6524]} \\
 &  & test doesn't add much value &  \\
 \cline{2-3}
 & Flaky Tests & flaky & \\
\hline
\end{tabular}
}
\end{center}
\end{table}

\begin{table}[t]
\caption{Extracted top n-gram keywords per project.}
\label{tb:keyword_project}
\begin{center}
\resizebox{\columnwidth}{!}{
\def\arraystretch{1.2}
\begin{tabular}{L{4.6cm}L{4.2cm}L{3.8cm}}
\hline
\textbf{Camel} & \textbf{Chromium} & \textbf{Gerrit} \\
\hline
\textbf{leak} & \textbf{leak} & \textbf{confusing} \\
\textbf{typo} & \textbf{flaky} & \textbf{typo} \\
\textbf{confusing} & \textbf{slow} & \textbf{flaky} \\
verbose & \textbf{unnecessary} & \textbf{unused} \\
deprecated & \textbf{simplify} & \textbf{\underline{bad}} \\
dead & \textbf{redundant} & \textbf{slow} \\
\textbf{slow} & \textbf{typo} & truncated \\
\textbf{unnecessary} & truncated & unnecessarily \\
document this & \textbf{\underline{ugly}} & not implemented yet \\
avoid & not implemented & \textbf{leak} \\
todo & \textbf{unused} & misleading \\
improve documentation & \textbf{\underline{bad}} & documentation \underline{is wrong} \\
complicated & \textbf{confusing} & coverage \\
remove \underline{ugly} warnings & odd & complicated \\
thread safe & the short term & performance degradation \\
reuse & \underline{clean up code} & documentation doesn't \\
missing & too verbose & undocumented \\
\underline{rid of} & \textbf{expensive} & \textbf{\underline{ugly}} \\
improve exception message if failed & isn't implemented & reword documentation \\
improve performance & too much & ambiguous \\
\hline
\end{tabular}
}
\resizebox{\columnwidth}{!}{
\def\arraystretch{1.2}
\begin{tabular}{L{3cm}L{3cm}L{3cm}L{3cm}}
\textbf{Hadoop} & \textbf{HBase} & \textbf{Impala} & \textbf{Thrift} \\
\hline
\textbf{unnecessary} & \textbf{flaky} & \textbf{flaky} & \textbf{unused} \\
\textbf{unused} & \textbf{unused} & \textbf{slow} & \textbf{leak} \\
\textbf{typo} & nit & \textbf{unnecessary} & \textbf{unnecessary} \\
\textbf{redundant} & \textbf{typo} & coverage & \textbf{typo} \\
nit & \textbf{leak} & \textbf{confusing} & \textbf{redundant} \\
\textbf{leak} & \textbf{\underline{ugly}} & \textbf{simplify} & \textbf{confusing} \\
\textbf{slow} & \textbf{redundant} & misleading & \textbf{simplify} \\
\textbf{flaky} & \textbf{unnecessary} & excessive & \textbf{flaky} \\
readability & \textbf{confusing} & overhead & coverage \\
\underline{clean up code} & too much & avoid & thread-safe \\
complicated & \textbf{\underline{bad}} & \textbf{expensive} & spurious \\
spurious & \textbf{slow} & improve error message & \underline{inconsistent} \\
reuse & \textbf{expensive} & \textbf{redundant} & abstract \\
\textbf{\underline{bad}} & misleading & rework & redundancy \\
\textbf{\underline{ugly}} & avoid & thread-safe & \textbf{\underline{ugly}} \\
not used & \textbf{simplify} & reduce duplicate code & outdated \\
cover & overhead & readability & missing \\
\underline{rid of} & dead lock & difficult & performance regression \\
\textbf{expensive} & readability & wasted space & extra \\
thread-safe & \underline{rid of} & verbose & unstable \\
\hline
\end{tabular}
}
\end{center}
\end{table}

We also summarize the top keywords for each project.
\cref{tb:keyword_project} presents the top 20 keywords on each project.
We note that the keywords are highlighted in bold if the number of projects where the keywords occur is greater than or equal to four (i.e. it appears in over half of projects).
From the table, we observe that nine keywords (e.g., leak, confusing, and unnecessary) are shared by different projects frequently.
We also find two projects (i.e. Chromium and HBase), one from each issue tracking system, containing all these nine common keywords; this indicates that technical debt is admitted similarly in different issue tracking systems although each project has some unique keywords.

\begin{figure}[t]
  \centering
  \includegraphics[width=0.7\columnwidth]{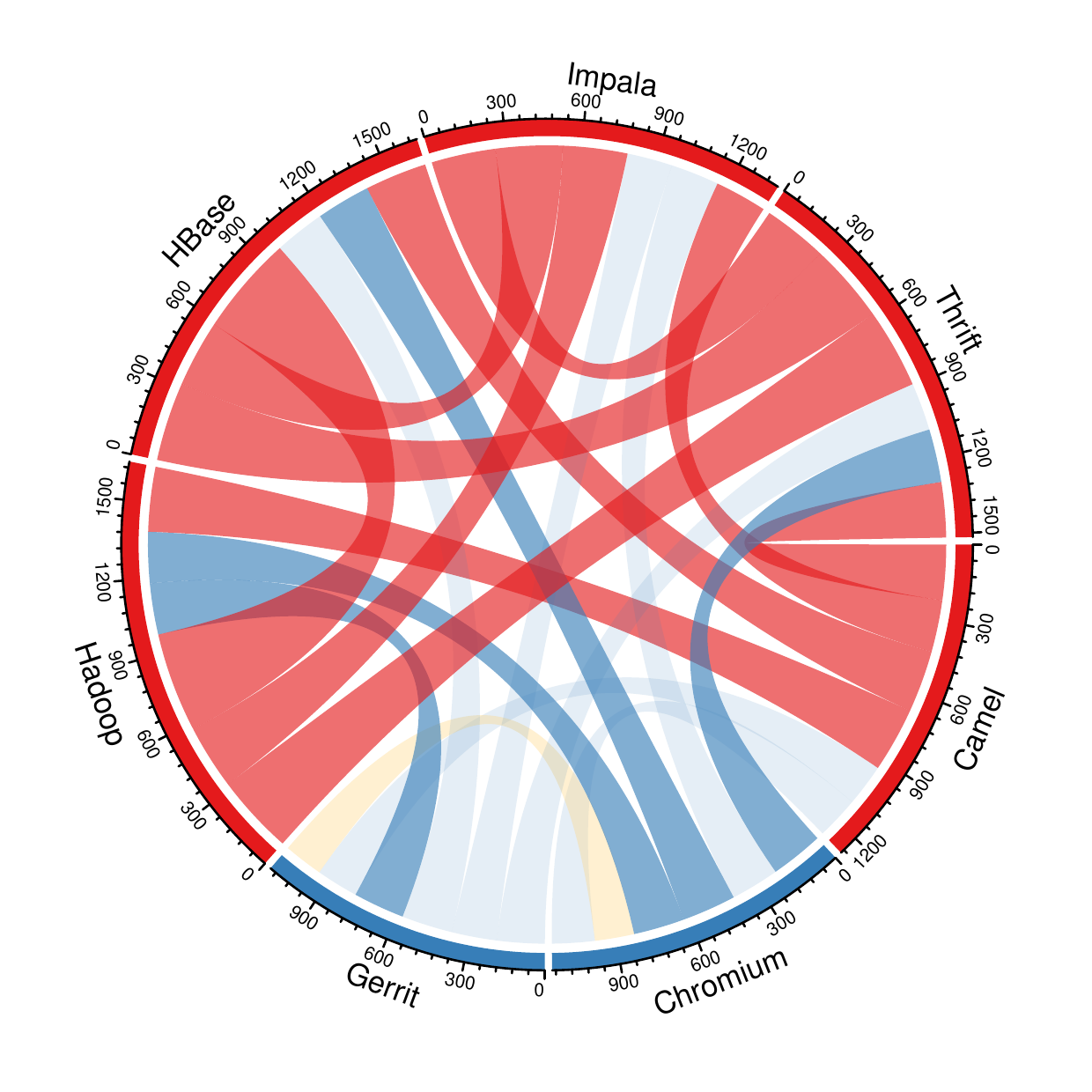}
  \caption{Relations (the number of common keywords) between different projects.}
  \label{f:number_of_shared_keywords}
  \vspace{-3mm}
\end{figure}

To gain a better understanding of how many keywords are shared by different projects, we first calculate the average number of extracted keywords from different projects and use the top 10\% (i.e., 2628) of the average number of most important keywords for analysis.
Then we plot a chord diagram \citep{gu2014circlize} illustrating the relations (i.e. the number of common keywords) between pairs of projects.

\cref{f:number_of_shared_keywords} shows the number of common keywords between different projects.
In the figure, the absolute value (i.e., the number of common keywords) measures the strength of the relation, and links are more transparent if the relations are weak (i.e., the number of common keywords is in the 30th percentile).
Moreover, the links between projects using Jira issue tracking system are colored red, and between projects using Google issue tracking system are colored yellow.
The links between the two issue tracking systems are colored blue.

We can see that although relations between pairs of projects seem equally strong, according to the sum of the number of shared keywords between projects, only Gerrit and Chromium share less than 1200 keywords with other projects; this indicates that these two Google projects have the weakest connection with others.
Besides, Chromium have strong relations with three projects using Jira.
Moreover, we can observe that the relation between the two Google projects is also weak, while there is no weak relation between projects using Jira.
This entails that there are discernible differences between SATD keywords in projects using Google Monorail (like Gerrit and Chromium) and Jira.
The projects using Jira might have more common ways to declare SATD compared to projects using Google Monorail.
This could be due to two potential reasons: 1) many developers in the Apache ecosystem work in multiple projects, so they use similar keywords across those projects; 2) the number of developers involved in the Google ecosystem is higher compared to the Apache ecosystem, so the variety of keywords is also higher.

In \cref{tb:n-grams,tb:keyword_project}, we can also observe that our keywords cover a few of the keywords from source code comments (see \cite{potdar2014exploratory}), such as `ugly', `inconsistent', `bad', `is wrong', `get rid of', and `clean up code', which are underlined in the table.
In addition to these common keywords, there are three more common keywords not listed in the table due to space limitation:  `hacky', `bail out', and `crap'.
Overall, we conclude that some keywords indicate SATD in both source code comments and issue tracking systems.

\begin{framed}
\noindent \textit{We find that extracted keywords are intuitive and potentially indicating types and indicators of SATD.
We also observe that although different projects share a great number of SATD keywords, projects using different issue tracking systems have less common keywords compared to projects using the same issue tracking system.
Source code comments and issue tracking systems have some common SATD keywords.
}
\end{framed}

\subsection{\textit{(RQ3) How Generic Is the Classification Approach Among Projects and Issue Tracking Systems?}} 
\label{sec:rq3}

In order to investigate the generalisability of our SATD detector over projects, we choose one project as the test project and the rest of the six projects as training projects using the configurations in \cref{sec:rq1.3}.
We then repeat this process for each project and calculate the average precision, recall, and F1-score over seven experiments.
We call this leave-one-out cross-project validation.
The results including F1-score, precision, and recall are presented in \cref{tb:cross_project_validation}.
We observe that, our approach achieves the average F1-score of 0.652, ranging between 0.561 to 0.709.
In comparison with the average F1-score achieved using stratified 10-fold cross-validation (i.e., 0.691) in \cref{sec:rq1.3}, the average F1-score decreases by a small margin (5.6\%) from 0.691 to 0.652.

\begin{table}[t]
\caption{Precision, recall, and F1-score when using six projects for training and the rest project for testing.}
\label{tb:cross_project_validation}
\begin{center}
\resizebox{0.7\columnwidth}{!}{
\def\arraystretch{1.2}
\begin{tabular}{@{\extracolsep{4pt}}C{2.5cm}C{1.5cm}C{1.5cm}C{1.5cm}@{}@{}}
\hline
\textbf{Project} & \textbf{Precision} & \textbf{Recall} & \textbf{F1-score} \\
\hline
\textbf{Camel} & 0.719 & 0.647 & 0.681 \\
\textbf{Hadoop} & 0.618 & 0.761 & 0.682 \\
\textbf{HBase} & 0.651 & 0.648 & 0.649 \\
\textbf{Impala} & 0.697 & 0.721 & 0.709 \\
\textbf{Thrift} & 0.693 & 0.668 & 0.679 \\
\textbf{Chromium} & 0.659 & 0.556 & 0.603 \\
\textbf{Gerrit} & 0.481 & 0.671 & 0.561 \\
\hline
\textbf{Avg.} & 0.645 & 0.667 & 0.652 \\
\hline
\end{tabular}
}
\end{center}
\end{table}

\begin{table}[b]
\caption{Precision, recall, and F1-score when using projects from the same issue tracking system for training and a projects in the other issue tracking system for testing.}
\label{tb:cross_issue_tracker_validation}
\begin{center}
\resizebox{\columnwidth}{!}{
\def\arraystretch{1.2}
\begin{tabular}{@{\extracolsep{4pt}}C{1.2cm}C{1.65cm}C{1.cm}C{1.1cm}C{1.cm}C{1.1cm}C{1.cm}C{1.1cm}@{}@{}}
\hline
\multirow{2}{*}{\textbf{\makecell{Issue\\ Tracker}}} & \multirow{2}{*}{\textbf{\makecell{Project\\(Target)}}} & \multicolumn{2}{c}{\textbf{Precision}} & \multicolumn{2}{c}{\textbf{Recall}} & \multicolumn{2}{c}{\textbf{F1-score}} \\
\cline{3-4}
\cline{5-6}
\cline{7-8}
 &  & Result & Diff. & Result & Diff. & Result & Diff. \\
\hline
 & & \multicolumn{6}{c}{Trained on projects using Google issue tracker} \\
\cline{2-8}
\multirow{5}{*}{\textbf{Jira}} & \textbf{Camel} & 0.553 & -23.0\% & 0.538 & -16.8\% & 0.545 & -19.9\% \\
 & \textbf{Hadoop} & 0.526 & -14.8\% & 0.611 & -19.7\% & 0.566 & -17.0\% \\
 & \textbf{HBase} & 0.506 & -22.2\% & 0.610 & -5.8\% & 0.553 & -14.7\% \\
 & \textbf{Impala} & 0.559 & -19.7\% & 0.619 & -14.1\% & 0.588 & -17.0\% \\
 & \textbf{Thrift} & 0.589 & -15.0\% & 0.590 & -11.6\% & 0.590 & \textbf{-13.1\%} \\
\cline{2-8}
 & \textbf{Avg.} & 0.546 & -18.9\% & 0.593 & -13.6\% & 0.568 & -16.3\% \\
\hline
 & & \multicolumn{6}{c}{Trained on projects using Jira issue tracker} \\
\cline{2-8}
\multirow{2}{*}{\textbf{Google}} & \textbf{Chromium} & 0.591 & -10.3\% & 0.537 & -3.4\% & 0.563 & -6.6\% \\
 & \textbf{Gerrit} & 0.488 & 1.4\% & 0.569 & -15.2\% & 0.526 & \textbf{-6.2\%} \\
\cline{2-8}
 & \textbf{Avg.} & 0.539 & -4.4\% & 0.553 & -9.3\% & 0.544 & -6.4\% \\
\hline
\end{tabular}
}
\end{center}
\end{table}

The second half of RQ3 concerns evaluating the generalizability of our CNN model (see \cref{sec:rq1.3}) over issue tracking systems.
We first choose one issue tracking system as the training system and the other one as the testing system.
We then use issue sections from the projects using the training system for training the model and issue sections from the other projects using the testing system for testing.
Then we switch the usage of two issue tracking systems.
We call this leave-one-out cross-issue-tracker validation.
\cref{tb:cross_issue_tracker_validation} shows the results (i.e., precision, recall, and F1-score).
The average precision, recall, and F1-score of projects are calculated and compared with the results obtained from leave-one-out cross-project validation (in \cref{tb:cross_project_validation}).
We find that the average F1-score achieved by the leave-one-out cross-issue-tracker validation is relatively worse compared to the leave-one-out cross-project validation: we achieve 6.4\% and 16.3\% decrease for models training on Jira and Google issue tracking systems respectively.

\begin{framed}
\noindent \textit{Our approach achieves the average F1-score of 0.652, ranging between 0.561 to 0.709, when applying leave-one-out cross-project validation. The average F1-score achieved by the leave-one-out cross-issue-tracker validation is declined by 6.4\% and 16.3\% for models training on Jira and Google issue tracking systems in comparison with leave-one-out cross-project validation.
}
\end{framed}

\subsection{\textit{(RQ4) How Much Data Is Needed for Training the Machine Learning Model to Accurately Identify Self-Admitted Technical Debt in Issues?}}
\label{sec:rq4}

\begin{figure}[b]
  \centering
  \includegraphics[width=0.8\columnwidth]{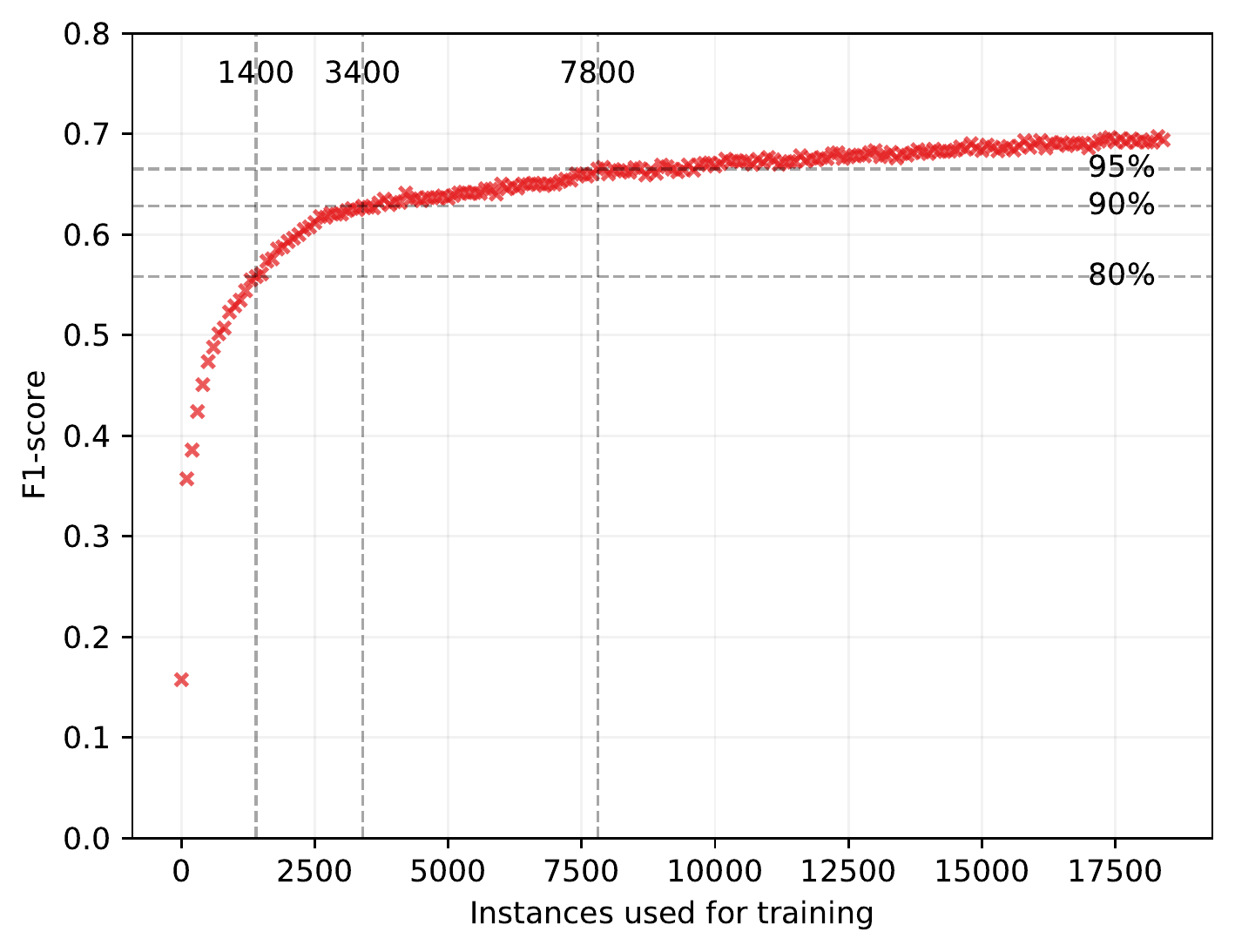}
  \caption{F1-score achieved by incrementally adding 100 issue sections into the training dataset.}
  \label{f:rq4}
  \vspace{-3mm}
\end{figure}

To answer this research question, we train our CNN model on datasets with a variety of sizes (number of issue sections).
More specifically, we first combine all the issue sections from different projects, then shuffle and split the combined dataset into ten equally-sized partitions for 10-fold cross-validation.
After that, we select one of the ten subsets for testing and the rest of the nine subsets for training.
Because we want to train the model on the datasets with a variety of sizes, we create an empty training dataset, add 100 issue sections to the training dataset each time, and train our model on the created training dataset.
Since we have $23,180$ issue sections in total, the subset for training contains $23180 \times \frac{9}{10} \approx 20862$ issue sections.
For each fold, we train the model on the dataset whose size increases from 100 to 20,862 at an interval of 100. 
Thus, we perform $\frac{20862}{100} + 1 \approx 209$ experiments for each fold.
This process is repeated ten times for each one of the ten folds.
In total, $10 \times 209 = 2090$ experiments are carried out and the average F1-score for different sizes of training dataset over ten folds is calculated.

The result of the average F1-score achieved while increasing the size of the training dataset is shown in \cref{f:rq4}.
We can observe that, while increasing the size of the training dataset from 100 to 2500 sections, the average F1-score goes up dramatically.
After the 2500 sections, the average F1-score improves slowly.
When the training dataset contains 20,862 sections, the highest F1-score (0.697) is achieved.
Besides, we find that in order to achieve 80\%, 90\%, or 95\% of the highest average F1-score, 1400, 3400, or 7800 sections are needed respectively (namely 6.7\%, 16.3\%, or 37.4\% of training data).

\begin{framed}
\noindent \textit{The average F1-score grows dramatically while increasing the size of the training dataset from 100 to 2500 sections.
After 2500 sections, the average F1-score improves slowly.
In order to achieve 80\%, 90\%, or 95\% of the highest average F1-score, 1400, 3400, or 7800 sections are needed respectively (namely 6.7\%, 16.3\%, or 37.4\% of training data).
}
\end{framed}

\section{Discussion}
\label{sec:discussion}

\subsection{Differences Between Identifying SATD in Source Code Comments and Issue Tracking Systems}

In recent years, a number of studies have investigated training machine learning models to automatically identify SATD in source code comments \citep{da2017using,flisar2019identification,ren2019neural}.
In contrast, our study focuses on SATD in issue tracking systems.
It is important to investigate the differences between identifying SATD in these two sources, because it could help us better understand the nature of SATD and build machine learning models to more accurately identify it in these and other sources.
Thus, in the following, we compare and discuss the differences between identifying SATD in code comments and issues.
Based on our experience gained during the manual issue analysis, we conjecture two major reasons for the difference among issues and source code comments, as explained in the following paragraphs.

\begin{table}[b]
\caption{Statistics for the source code comment dataset and issue tracking system dataset.}
\label{tb:statistics_datasets}
    \begin{center}
    \resizebox{\columnwidth}{!}{
    \def\arraystretch{1.2}
        \begin{tabular}{C{3cm}C{4.2cm}C{2.cm}C{2.3cm}}
        \hline
        \textbf{Source} & \textbf{Avg. Length of Issue Sections / Code Comments} & \textbf{\# of Sections / Comments} & \textbf{Vocabulary Size} \\
        \hline
        Source code comments & 10.9 & 62275 & 31728 \\
        Issue tracking systems & 35.4 & 23180 & 37202 \\
        \hline
        \end{tabular}
    }
    \end{center}
\end{table}

First, \textbf{the diversity of issues is much higher than source code comments.}
In the work by \cite{steidl2013quality}, source code comments are categorized into seven types, namely \textit{code, copyright, header, member, inline, section,} and \textit{task comments}.
The first type, \textit{code comments}, refer to code that is commented out; this certainly does not indicate technical debt.
The types \textit{copyright, header, member, inline,} and \textit{section comments} are descriptive comments, which have a small chance of indicating technical debt.
The only type that mostly concerns SATD is \textit{task comments}, as these are notes left by developers indicating code that needs to be implemented, refactored, or fixed.

In contrast, the scope of information stored in issue tracking systems is much broader as compared to code comments.
\cite{merten2015requirements} categorized issue sections into 12 types, namely \textit{issue description, request, issue management, scheduling, implementation proposal, implementation status, clarification, technical information, rationale, social interaction, spam,} and \textit{others}. 
Apart from \textit{issue management, scheduling, social interaction,} and \textit{spam}, all other 8 types of issue sections could potentially indicate technical debt.
Because the diversity of issue types is much higher than source code comments, it is harder for machine learning models to accurately capture SATD in issue types than in source code comments.
Furthermore, we present a comparison between the key statistics for the two datasets in \cref{tb:statistics_datasets}.
We observe that the average length of issue sections is much longer than source code comments; this indicates that issue sections contain more information than source code comments in general.
Furthermore, while there are three times more source code comments than issue sections in the two datasets, the vocabulary size of issue sections is a little higher. 
These statistics confirm that the diversity of issues is higher than source code comments.  

Second, \textbf{some types of SATD are different in source code comments and issue tracking systems.}
In the following paragraphs, we highlight two specific types of SATD, namely \textit{defect debt} and \textit{requirement debt}, that have key differences in code comments and issues.
In the case of defect debt, only the defects that are left unresolved, can be classified as defect debt (see \cref{tb:statistics_type_projects}).
For source code, \textit{task comments} denoting that bugs need to be fixed can be directly classified as \textit{defect debt}, since they indicate that bugs are reported to be fixed at a later stage.
An example is shown below:

\begin{displayquote}
    \textit{``TODO: may not work on all OSes''} - [Defect debt from JMeter code comments]
\end{displayquote}

However, in issue tracking systems, most bugs are reported and resolved immediately.
Thus, in order to capture defect debt, we need to first identify defects and then judge whether fixing them is postponed or not.
Only if fixing the defects is deferred or ignored, they can be tagged as \textit{defect debt}.
To exemplify this distinction, we show an example of an issue section that denotes defect debt and one that does not:

\begin{displayquote}
    \textit{``I do not think it is a critical bug. Deferring it to 0.14.''} - [Defect debt from Hadoop issues]

    \textit{``...thank you for reporting the bug and contributing a patch.''} - [Non-defect debt from Hadoop issues]
\end{displayquote}

Similarly, \textit{requirement debt} also manifests differently in source code comments and issues.
According to the definition of requirement debt (see \cref{tb:statistics_type_projects}), it reflects partially implemented functional or non-functional requirements.
In source code comments, if the requirement is not completely satisfied, developers might leave a code comment, such as:

\begin{displayquote}
    \textit{``TODO support multiple signers''} - [requirement debt from JMeter code comments]
    
\end{displayquote}

The above comment can be simply annotated as \textit{requirement debt}, as we know that the requirement is only partially implemented.
However, in issue tracking systems, things are different.
As mentioned above, \textit{implementation proposal} is a type of issue section, that expresses new requirements. 
But such requirements may be partially or fully implemented in the meantime; that can be inferred through the ensuing discussion in the issue.
To identify \textit{requirement debt} in issues, it is necessary to differentiate partially implemented requirements from requirements that are fully implemented or not at all.
The following examples show an issue section that denotes a requirement debt and one that does not:

\begin{displayquote}
    \textit{``The backend (in master) has everything in place now to support this, but the frontend still needs to be adapted.''} - [Requirement debt from Gerrit issues]

    \textit{``It would be good to add a script to launch Hive using Ranger authorization.''} - [Non-requirement debt from Impala issues]
\end{displayquote}

Although identifying SATD in issue tracking systems is harder than in source code comments, it does not require a much larger amount of data to achieve decent accuracy when training machine learning models.
Compared with previous work by \cite{da2017using} that identifies SATD in source code comments, their incremental training curves are similar to ours (see \cref{f:rq4}).
More specifically, the F1-score increases dramatically when the training dataset is small.
After that, it goes up moderately. 
In order to achieve 90\% of the accuracy of SATD identification, our model needs 16.3\% (3400) issue sections on the issue SATD dataset, while about 23.0\% (11800) source code comments are required on the source code comment SATD dataset. 
Therefore, relatively small datasets can achieve decent accuracy on both code comments and issue sections.


\subsection{Similarity Between SATD Keywords Extracted from Source Code Comments and Issue Tracking Systems}

\begin{table}[b]
\caption{Top 10 n-gram SATD keywords from issue tracking systems and source code comments.}
\label{tb:keyword-comparison}
\begin{center}
\resizebox{\columnwidth}{!}{
\def\arraystretch{1.2}
\begin{tabular}{L{4cm}L{3.2cm}L{3.8cm}}
\hline
\textbf{\makecell[l]{Unique Keyword \\(Issue Tracking Systems)}} & \textbf{Common Keyword} & \textbf{\makecell[l]{Unique Keyword \\(Source Code Comments)}} \\
\hline
performance & why & todo \\
clean & improve & fixme \\
typo & leak & hack \\
remove & probably & should \\
flaky & perhaps & workaround \\
unused & better & defer argument checking \\
slow & instead & xxx \\
refactor & wrong & bug \\
warnings & missing & not needed \\
confusing & deprecated & implement \\
\hline
\end{tabular}
}
\end{center}
\end{table}

\begin{figure}[t]
  \centering
  \includegraphics[width=0.8\columnwidth]{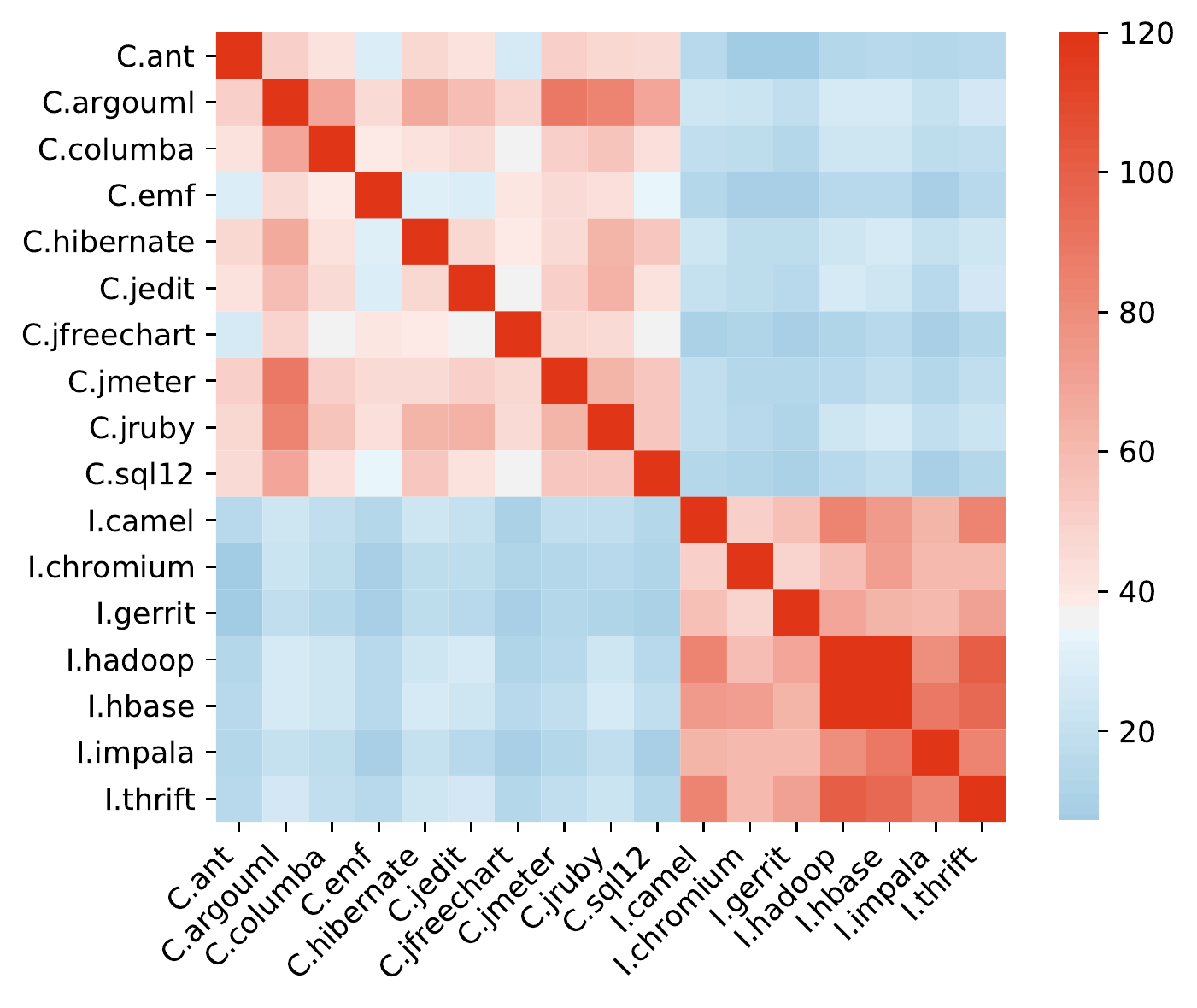}
  \caption{Number of common keywords between different projects.}
  \label{f:number_of_shared_keywords_all}
\end{figure}

In \cref{tb:n-grams} and \cref{tb:keyword_project}, some keywords are underlined to highlight common SATD keywords in both issues and code comments.
We observe that there are some common SATD keywords shared between source code comments and issues, such as `ugly', `bad', and `get rid of'.
In order to get a deeper understanding of the differences between SATD keywords from the two sources, we extract the top 10\% (i.e., 272) unigram to trigram SATD keywords (across all projects) from source code comments and issue tracking systems.
Subsequently, we compare the top 272 extracted SATD keywords from the two sources.

The top unique and common keywords are presented in \cref{tb:keyword-comparison}, while the heatmap in \cref{f:number_of_shared_keywords_all} illustrates the number of common keywords between the two sources.
In \cref{f:number_of_shared_keywords_all}, the symbol attached before the project names (i.e., \textit{C} or \textit{I}) refers to the data source (i.e., \textit{code comments} or \textit{issues}).
We observe that there are fewer keywords shared between issues and code comments compared to keywords shared between issues or between code comments.
More specifically, we find on average that 73.4 keywords are shared between projects in issues, 48.4 keywords are shared between projects in code comments, and only 16.8 keywords are shared between projects across the two sources.
The finding indicates that \textbf{there is a certain similarity between SATD in issues and code comments, but the similarity is limited}.

\subsection{Implications for Researchers and Practitioners}

Based on the findings, we suggest the following directions for researchers:

\begin{itemize}
    \item Our work provides a deep learning approach to automatically identify SATD from issue tracking systems.
    The proposed approach can enable researchers to automatically identify SATD within issues and conduct studies on the measurement, prioritization, as well as repayment of SATD in issue tracking systems on a large scale.
    
    \item To enable further research in this area, we make our issue SATD dataset publicly available\cref{l:data}.
    The dataset contains 23,180 issue sections, in which 3,277 issue sections are classified as SATD issue sections.
    
    \item We found that relatively small datasets can achieve decent accuracy in identifying SATD on both source code comments and issues. 
    We thus recommend that researchers explore SATD in other sources (i.e., pull requests and commit messages) and contribute a moderate-sized dataset for automatic SATD identification in the corresponding sources.
    
    \item Our findings suggest that there is a certain similarity between SATD in issues and in source code comments, but the similarity is limited.
    We encourage researchers to study the differences between SATD in these and other sources, e.g. in pull requests or commit messages. 
    This could advance the understanding of SATD in the different sources.
    
    \item 
    Although our study experimented with the generalizability of our approach across projects and across issue tracking systems, the scope of our study is still limited.
    Thus, we recommend that researchers investigate the applicability of our approach to other projects (esp. industrial projects) and other issue tracking systems. 
    If possible, we advise them to make their datasets publicly available to be used for training new SATD detectors.
    
    \item Because of the high diversity of issues and the different forms of SATD in issues, SATD identification within issues is harder than in source code comments.
    However, further research can potentially improve the F1-score obtained in our study (e.g., through using other machine learning techniques or trying richer datasets in the software engineering domain for transfer learning).
    
\end{itemize}

We also propose a number of implications for software practitioners:

\begin{itemize}
    \item Our SATD identification approach can help software developers and especially project managers to evaluate the quality of their project.
    For instance, project managers can use this tool to track SATD in issue tracking systems along evolution. 
    If the accumulated SATD reaches a threshold, then more effort may need to be spent in paying it back.
    
    \item We recommend that tool developers use our SATD identifier in their toolsets and dashboards and experiment with them in practice.
    
    \item We encourage practitioners to study carefully the SATD keywords listed in our results. 
    This will help them to understand in practice the nature of SATD, how to better formulate it themselves and how to recognize SATD stated from others.
    
    \item Our findings can help practitioners better understand the differences between SATD in different sources, e.g., defect debt is identified differently in source code comments compared to issue tracking systems.
    This can also help practitioners better identify SATD in different sources.
\end{itemize}

\section{Threats to Validity}
\label{sec:validity}

\subsection{Threats to Construct Validity}

Construct validity reflects the correctness of operational measures for the studied subjects.
We observed that only a small amount of issue sections are classified as SATD issue sections.
To accurately measure the accuracy of machine learning models on SATD identification, we chose precision, recall, and F1-score as evaluation metrics.
These metrics have been used in previous similar studies \citep{da2017using,huang2018identifying}, and are well-established for this type of work.

\subsection{Threats to External Validity}

Threats to external validity concern the generalizability of our findings.
Because we trained machine learning models on our issue SATD dataset, 
the data selected and analyzed might influence the generalizability of our findings.
In order to partially mitigate this threat, we randomly selected seven open-source projects using two different issue tracking systems, being maintained by mature communities, and containing sufficient issue data.
Besides, to ensure the collected issues are sufficiently representative of issues in each project, we calculated the size of the statistically significant sample based on the number of issues in each project.
We then randomly selected issues according to the size of the statistically significant sample.
In \cref{sec:rq4}, we showed that the size of our dataset is sufficient for training machine learning models to identify SATD from issue tracking systems.
When evaluating the predictive performance of machine learning models, we use stratified 10-fold cross-validation to mitigate the bias caused by random sampling.
Moreover, in \cref{sec:rq3} we evaluated the generalizability of our approach across projects and across issue tracking systems.

However, because of the nature of open-source projects, developers tend to communicate online through tools, such as issue tracking systems and mailing lists; this facilitates new contributors to understand the details of issues and contribute to projects.
In contrast, industrial projects have most developers working at the same premises; so, they tend to efficiently communicate the details of issues offline. 
This limits the generalizability of our results to such projects.
In conclusion, our findings may be generalized to other open-source projects of similar size and complexity and of similar ecosystems.

\subsection{Threats to Reliability}

Reliability considers the bias that researchers may induce in data collection and analysis.
Moreover, we manually classified issue sections as different types of SATD or not.
To reduce this bias, the issue data was first analyzed manually by four independent researchers.
Then the first author analyzed issue samples and calculated Cohen's kappa coefficient between his output and that of the independent researchers.
If the agreement was good (i.e., Cohen's kappa coefficient is above 0.7), the classification is considered complete.
If not, the first author discussed the classification differences between them to reach a consensus.
Subsequently, they improved the classification, and Cohen's kappa coefficient was calculated again to ensure the level of agreement was good.

Furthermore, our results depend on the data analysis methods we use.
In terms of machine learning approaches, we have chosen some of the most commonly approaches used by researchers and practitioners.
Besides, we follow established guidelines for tuning hyper-parameters \citep{zhang2017sensitivity} and exploring transfer learning \citep{semwal2018practitioners}. Finally, and most importantly, we make our issue SATD dataset publicly available\cref{l:data} in the replication package.

\section{Conclusion}
\label{sec:conclusion}

In this work, we investigated SATD identification with respect to \textit{accuracy}, \textit{explainability}, and \textit{generalizability} in issue tracking systems.
We contributed a dataset including 23,180 issue sections classified as SATD sections or non-SATD sections from seven open-source projects using two issue tracking systems.
Moreover, we compared different machine learning algorithms and propose a CNN-based approach to identify SATD in issues with an F1-score of 0.686.
Furthermore, we explored the effectiveness of transfer learning using other datasets to improve the F1-score of SATD identification from 0.686 to 0.691.
In addition, we identified a list of n-gram top SATD keywords, which are intuitive and can potentially indicate types and indicators of SATD.
Besides, we observed that projects using different issue tracking systems have less common SATD keywords compared to projects using the same issue tracking system.
We also evaluated the generalizability of our approach. 
The results show our approach achieves the average F1-score of 0.652, ranging between 0.561 to 0.709, using leave-one-out cross-project validation; when applying leave-one-out cross-issue-tracker validation, the average F1-score is dropped by 6.4\% and 16.3\% for models training on Jira and Google issue trackers compared to using leave-one-out cross-project validation.
Finally, we investigated the amount of data needed for our approach.
We showed that only a small amount of training data is needed to achieve good accuracy.

In the future, we tend to explore the differences between SATD in different sources.
We also aim to use the ensemble learning technique to improve the predictive performance of machine learning models by combining different classifiers.
Moreover, we plan to further analyze our dataset to identify different types (or indicators) of SATD in issue tracking systems.

\bibliographystyle{spbasic}      
\bibliography{main}


\end{document}